\documentclass[aps,10pt,superscriptaddress,amsfonts, amssymb,amsmath,twocolumn]{revtex4-1}

\usepackage{graphicx}
\usepackage{mathtools}
\usepackage{bm}
\usepackage{bigints}
\usepackage{booktabs}
\usepackage{makecell}
\usepackage{xr}
\usepackage{hyperref}
\usepackage{mathbbol}  
\usepackage{xcolor}
\usepackage{geometry}
\newcommand\figref{Fig.~\ref}
\newcommand{\tabref}[1]{Table \ref{#1}}
\newcommand{\norm}[1]{\left\lVert#1\right\rVert}

\newcommand{\tb}[1]{\textcolor{black}{#1}}

\begin{document}

\title{Base flow decomposition for complex moving objects in linear hydrodynamics: application to helix-shaped flagellated microswimmers}

\author{Ji Zhang}
\affiliation{Beijing Computational Science Research Center, Beijing 100193, China}
\email{zhangji@csrc.ac.cn}
\author{Mauro Chinappi}
\affiliation{Department of Industrial Engineering,
  University of Rome, Tor Vergata,
  Via del Politecnico 1, 00133 Roma, Italia.}
\email{mauro.chinappi@uniroma2.it}
\author{Luca Biferale}
\affiliation{Dept. of Physics and INFN,
  University of Rome, Tor Vergata,
  Via della Ricerca Scientifica 1, 00133 Roma, Italia.}
\email{Luca.Biferale@roma2.infn.it}

\date{\today}

\begin{abstract}
The motion of microswimmers in complex 
flows is ruled by the interplay between swimmer propulsion 
and the dynamics induced by the fluid velocity field.
Here we study the motion of a chiral microswimmer whose propulsion 
is provided by the spinning of a helical tail with respect to its body in
a simple shear flow. 
Thanks to an efficient computational strategy that allowed us to 
simulate thousands of different trajectories, we show that the tail
shape dramatically affects the swimmer's motion.
In the shear dominated regime, 
the swimmers carrying an elliptical helical tail
show several different Jeffery-like (tumbling) trajectories depending
on  their initial configuration.
As the propulsion torque increases, a progressive regularization
of the motion is observed until, in the propulsion dominated regime,
the swimmers converge to the same final trajectory independently on the initial configuration.
Overall, our results show that elliptical helix swimmer presents a much richer variety of trajectories
with respect to the usually studied circular helix tails.
\end{abstract}


\maketitle

\section{Introduction}

Several microorganisms move in liquids thanks to rotating flagella.
For instance, the bacterium {\sl Escherichia coli} 
has several flagella that form a rotating helical bundle~\cite{berg2008coli},
while other bacteria, like
{\sl Pseudomonas aeruginosa}, exploit the same propulsion strategy but 
using a single helical flagellum~\cite{qian2013bacterial,sartori2018wall}.
The high swimming speed and the relatively simple geometry 
of such a kind of chiral microswimmers make them suitable
for various applications and, in the last decade, artificial 
versions of flagellated microswimmers have been proposed 
for micromanipulation~\cite{zhang2010artificial} 
and drug delivery~\cite{mhanna2014artificial}.

The interaction between helical flagellated microswimmers 
and the environment presents a rich behavior that has received 
extensive attention in the past decades~\cite{lauga2009hydrodynamics,elgeti2015physics}.
Close to interfaces, helical flagellated microswimmers 
follow circular trajectories that are clockwise for
solid walls~\cite{lauga2006swimming,guccione2017diffusivity,shum2010modelling}
and \tb{counterclockwise for
liquid-air interfaces~\cite{di2011swimming,pimponi2016hydrodynamics,hu2015physical,bianchi20193d}}.
Far from the wall, the hydrodynamic of active microorganisms is 
highly affected by the local flow conditions.
A relevant phenomenon is rheotaxis, 
i.e. the movement resulting from
fluid velocity gradients.
As shown by Fu et al.~\cite{fu2012bacterial}, the rheotaxis of flagellated 
microswimmers with helical tail is a purely physical
phenomenon due to interplay between velocity gradients and the shape of chiral flagella. 
Indeed, for a passive helix, the shear induces Jeffery-like tumbling motion parallel  
to the  shear plane~\cite{fu2009separation}. 
Along the orbit, elongated helices spend more time aligned with streamlines.
Since this configuration is not symmetric
with respect to the shear plane, a chirality-dependent drift generally sets in.
For active helical microswimmer, the passive chirality-induced drift
is often overwhelmed by the propulsion: the shear results in 
a preferential orientation of the swimmers along which,
thanks to the self-propulsion, the swimmer moves~\cite{fu2012bacterial,rusconi2014bacterial}. 
Hence, the swimming direction is ruled  by the shear,
likely preventing the possibility of controlling the orientation of microswimmers 
in an assigned flow~\cite{fu2012bacterial}. 

A way to escape from the monotonous rheotaxis in shear flow is to 
increase the number of degrees of freedom (DOF) of the microswimmers,
for instance employing multiple tails~\cite{kanehl2014fluid}, 
or adaptively changing the angle between body and tail(s)~\cite{riley2018swimming}. 
The existence of external flexibility, however, complicates the control of microswimmers particularly in view of possible technological applications. 
Another possibility to escape from the rheotaxis is to break some symmetries of the swimmer geometry. 
In this aspect, interestingly, it has recently been shown that 
the change of the cross-section of the ellipsoids from circle to ellipse can lead 
to chaotic 
orbits~\cite{einarsson2016tumbling, einarsson2015angular, thorp2019motion}.

Inspired by this phenomenon, we numerically analyzed the dynamics
of a microswimmer made by an axisymmetric body and by an elliptical helix, 
(i.e. a helix that lies on an elliptical cylinder) in a shear flow.
The possible presence of a large variety of different trajectories,
requires a systematic exploration of a large number of initial conditions.
This, in turn,  pushed us to develop and apply  
an accurate and fast computational approach  
based on a decomposition of dynamics in an active and a passive motion
that allowed to speed-up the simulations and to collect, for
each case, thousands of trajectories. 
Our results show that the elliptical helix swimmer
presents a much richer variety of possible 
trajectories with respect to the well studied
circular helix tails. In particular, we found 
for an elliptical helical tail 
a much higher spinning frequency is needed to control the asymptotic swimming regime.

%%%%%%%%%%%%%%%%%%%%%%%%%%%%%%%%%%%%%%%%%%%%%%%%%%%%%%%%%%%%%%%%
\section{Set-up and Method} 
\label{sct_method}

%%%%%%%%%%%%%%%%%%%%%%%%%%%%%%%%%%%%%%%%%%%%%%%%%%%%%%%%%%%%%%%%
\begin{figure}%[htb]
%  \centering
  \includegraphics[width=\columnwidth]{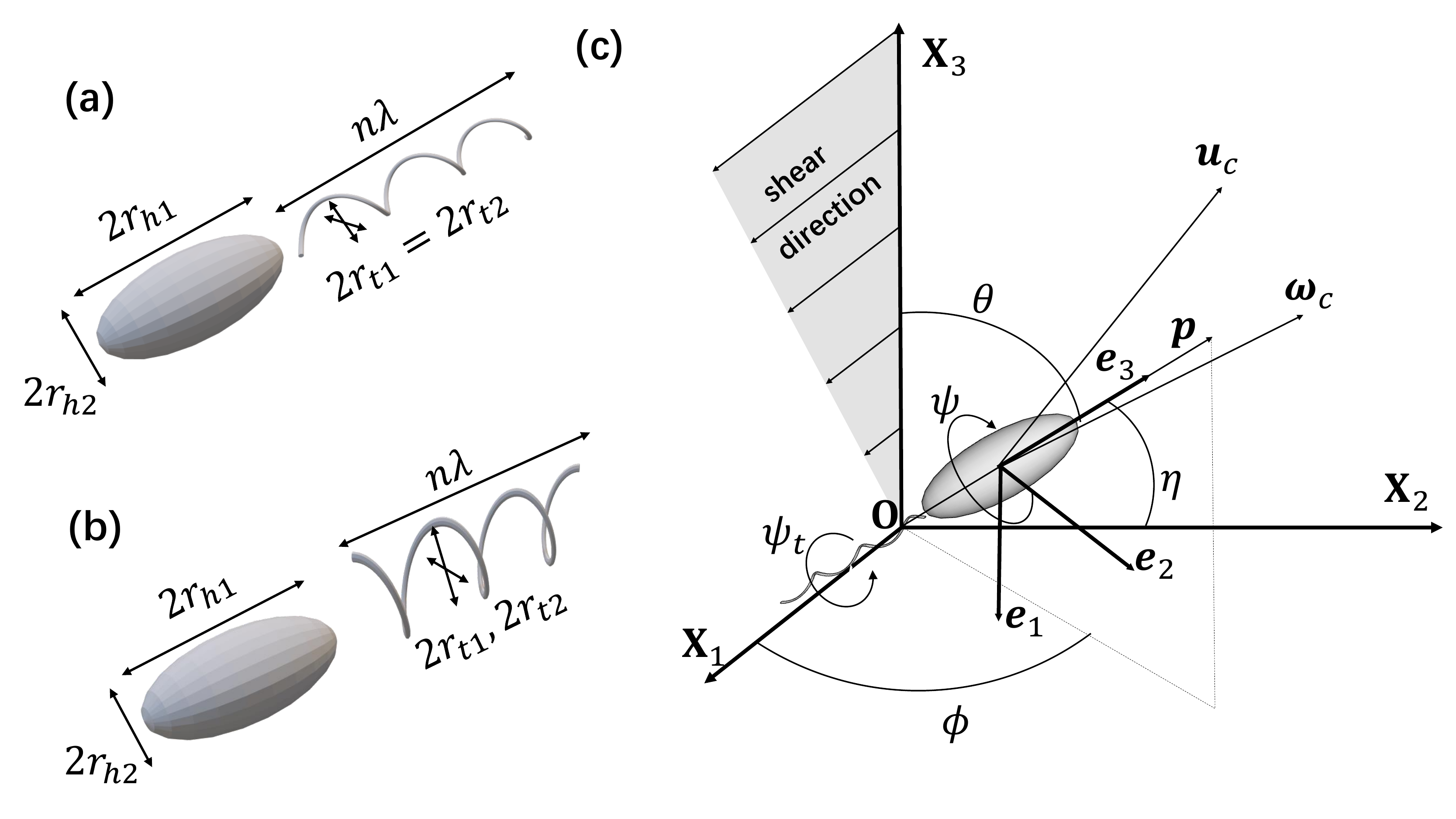}
  \caption{Sketch of microswimmer locomotion in a shear flow. 
  The microswimmer body is a prolate ellipsoid of major axis $r_{h1}$ and minor axis $r_{h2}$.
  In model I (a), the tail is a circular helix, i.e. a helix built on a circular cylinder of
  radius $r_{t1}$, while in model II (b), the tail is an elliptical helix, i.e.
  a helix built on an elliptical cylinder of radii $r_{t1}$ and $r_{t2}= 3 r_{t1}$.
  (c) The shear flow is in the $\textbf{X}_1\textbf{X}_3$ plane of the global coordinate 
  frame $\textbf{O}\textbf{X}_1\textbf{X}_2\textbf{X}_3$. 
  A body coordinate frame $\textbf{O}\bm{'}\textbf{e}_1\textbf{e}_2\textbf{e}_3$ moves with the
  swimmer body. 
  The polar $\theta$ azimuthal  $\phi$ and rotation $\psi$ angles are used to 
  describe the orientation of the body frame with respect to the global frame.
  Moreover, we also define the angle between the swimmer axis $\bm{p} = \textbf{e}_3$ 
  and $\textbf{X}_2$ as $\eta = \arccos (\sin \theta \sin \phi)$.
  The tail rotates with respect to the $\textbf{e}_3$ so that
  each point of the rigid tail describes a circle in the plane $\textbf{e}_2\textbf{e}_3$, 
  with $\psi_t$ the corresponding rotation angle.
  The motion of the microswimmer body is completely defined when 
  the translational velocity $\bm U$  of the body center, the 
  body rotational velocity $\bm \Omega$ and the tail spinning
  $\Omega_t = \dot{\psi_t}$ are given.
  \label{fig_setup}
  }
\end{figure}
%%%%%%%%%%%%%%%%%%%%%%%%%%%%%%%%%%%%%%%%%%%%%%%%%%%%%%%%%%%%%%%%

Two kinds of microswimmers are compared in this study:
one with a circular helical tail, 
i.e. a helix that lies on a circular
cylinder (model I) 
and one with an elliptical helical tail, 
i.e. a helix that lies on a 
cylinder of elliptical section (model II), see~\figref{fig_setup}.
For both models, the body is a prolate 
ellipsoid of radii $r_{h1}$ and $r_{h2}$, the center of the 
body is indicated as $\bm x_c$. 
A body coordinate frame $\textbf{O}\bm{'}\textbf{e}_1\textbf{e}_2\textbf{e}_3$ 
with origin at $\bm x_c$ and $\textbf{e}_3$ oriented as the 
major ellipsoid axis is defined. 
Concerning the tail, its centerline follows the helix equation 
in the body coordinate frame
\begin{align}
  {\bm r}= (r_{t1} \cos(2 \pi s), r_{t2} \sin(2 \pi s), \lambda s - \delta_{bt}) \, , 
\end{align}
where $s \in [-n/2, n/2]$ with $n$ the number of periods of the tail, 
$\delta_{bt}$ is the distance from $\bm x_c$ to the tail center, 
$\lambda$ is the pitch of the helix
and $r_{t1}$ and $r_{t2}$ are the radius of the elliptical cylinder
on which the helix lies.
For circular helix, $r_{t1}=r_{t2}$, while for elliptical
helix, $r_{t1} = 3 r_{t2}$.
The flagellum section is a cylinder of radius $\rho_t$. 
All the details of the swimmer geometry are reported in the 
appendix~\ref{sec_apx_method}.

%%%%%%%%%%%%%%%%%%%%%%%%%%%%%%%%%%%%%%%%%%%%%%%%%%%%%%%%%%%%%%%%%%%
\begin{figure*}
    \includegraphics[width=\textwidth]{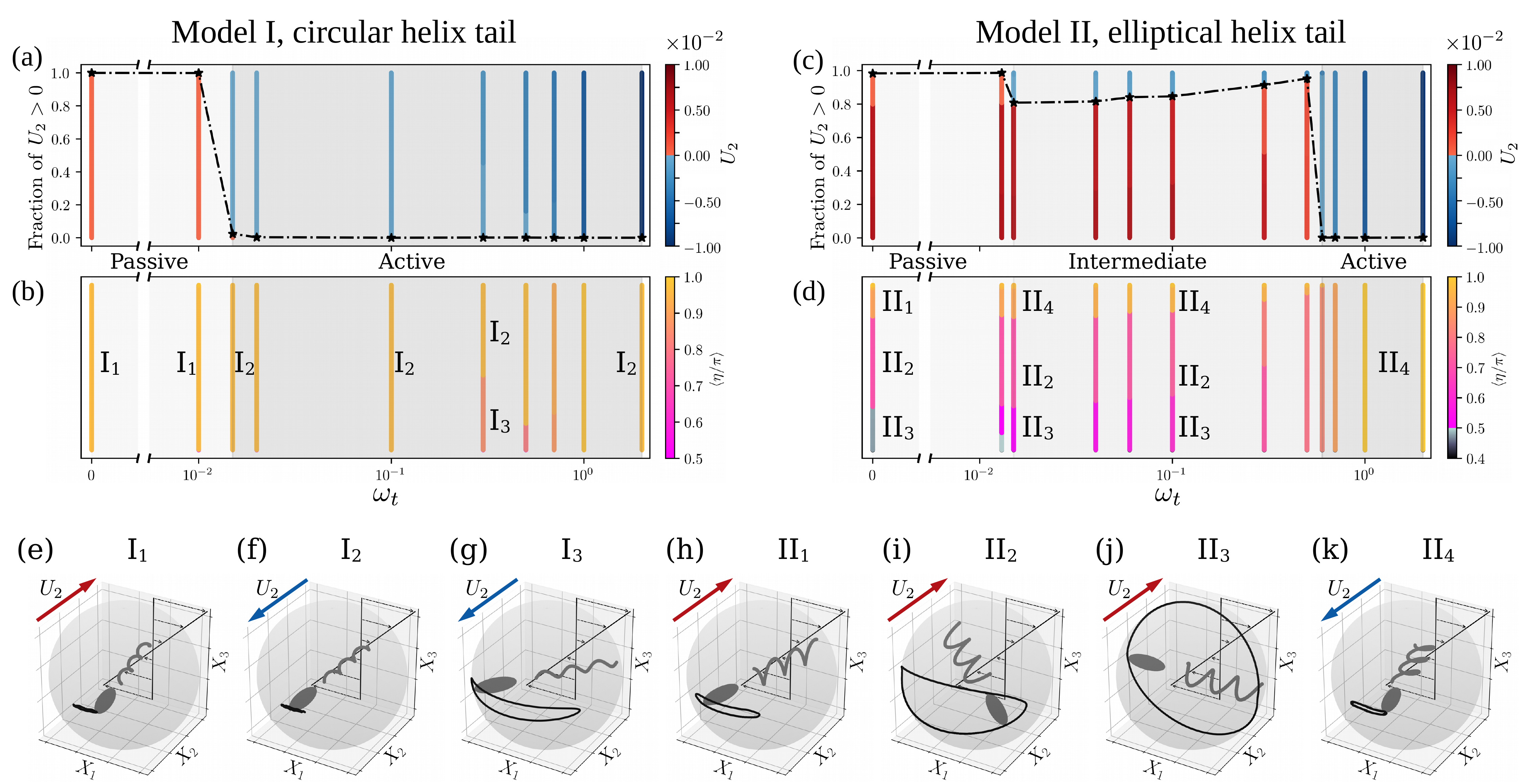}
  \caption{
Microswimmer motion in a simple shear flow. 
For each case, from $10^3$ to $10^4$ simulations with different initial conditions were
run. Panel a) reports the fraction of the circular helix swimmers
that have a drift velocity oriented as $\textbf{X}_2$ ($U_2 > 0$).
The bars in panel a) indicate the lateral velocity $U_2$ for the different initial conditions
while the bars in panel b) refer to the normalized average angle 
$\langle \eta / \pi \rangle$ between $\textbf{X}_2$ and 
the microswimmer head orientation $\bm p$.
In the passive regime, all the trajectories
converge to the same final state ($\mathrm{I_1}$, panel e) where the  
swimmer is oriented as $-\textbf{X}_2$ while its velocity is 
$U_2 \cong 10^{-4}$.
In the active regime, the swimmer is again oriented as $-\textbf{X}_2$, 
but $U_2 <0$, see configuration $\mathrm{I_{2}}$ (f) 
and $\mathrm{I_{3}}$ (g). 
The dot-dashed line corresponds to the fraction of swimmers  
for which $U_2 > 0$.
Panel $(c)$ and $(d)$ refer to velocity $U_2$ and orientation $\langle \eta / \pi \rangle$
for  elliptical helix tail. 
Here, a new intermediate regime appears between the active and
passive regime. In this intermediate regime,
both positive and negative drift velocities $U_2$ are possible. 
Panels (e)-(k) report examples of the swimmer \tb{Jeffery-like tumbling} 
motion \tb{(shear in the $X1X3$ plane)}.
The solid lines on the spheres represent the direction of 
the swimmer axis $\bm p$ along one period while the
red and blue arrows refer to the direction of the average velocity along $X_2$.
 \label{fig_phase_map}
}
\end{figure*}
%%%%%%%%%%%%%%%%%%%%%%%%%%%%%%%%%%%%%%%%%%%%%%%%%%%%%%%%%%%%%%%%%%%

The microswimmer has 7 degree of freedoms (DOFs): 
3 translation DOFs $\bm x_c = (x_{c1},x_{c2},x_{c3})$, 
3 rotational DOFs $\bm \theta_c = (\theta, \phi, \psi)$ 
plus the tail orientation $\phi_t$ with respect to the body. 
The body orientation is defined by the unit vector
$\bm p=(\sin \theta \cos \phi, \sin \theta \sin \phi, \cos \theta)$ 
here expressed as a function of the polar $\theta$ and the azimuthal $\phi$ angles. 
It is also instrumental to define the angle $\eta = \arccos (\sin \theta \sin \phi)$ 
between $\bm p$ and $\textbf{X}_2$, see~\figref{fig_setup}. 
The value $\eta = 0$ ($\eta = \pi$) corresponds to a configuration where 
the microswimmer is perpendicular 
to the shear plane and points toward positive (negative) $\textbf{X}_2$, 
while $\eta = 0.5 \pi$ corresponds to the microswimmer 
lying in the shear plane.
The swimmer body moves with translational velocity $\bm U$ 
and rotational velocity $\bm \Omega$ 
while the tail spins at a constant speed and consequently $\Omega_t = \dot{\psi_t}$.

The governing equations of fluid velocity $\bm{u}$ and pressure $p$ 
fields are the Stokes equations
\begin{align}
    \nabla\cdot\bm{u} &= 0 \, , 
    \label{eq01}\\
    \mu\nabla^2 \bm{u} &= \nabla p \, , 
    \label{eq02}
\end{align}
with $\mu$ the fluid viscosity. 
The no-slip boundary condition is applied on the surfaces of the head
\begin{align}
  \bm u(\bm x) &= \bm U + \bm \Omega \times \bm r \, ,  
  \label{eq_bc_head}
\end{align}
and of the tail of the microswimmer
\begin{align}
  \bm u(\bm x) &= \bm U + (\bm \Omega + \Omega_t \bm p) \times \bm r \, , 
  \label{eq_bc_tail}  
\end{align}
where in both equations $\bm r$ indicates the relative position of 
the boundary point with respect to the center of the swimmer head $\bm x_c$.
Note that, in general, $\bm U$ and $\bm \Omega$ 
are not parallel to the swimmer orientation $\bm p$.
Thus, there exists no simple relation between the active spin $\Omega_t$ and the velocities 
$(\bm U, \bm \Omega)$.

The method for the solution of the swimming problem 
is briefly sketched in the following while details are reported in the appendix~\ref{sec_apx_method}.  
The fundamental step is to get the swimmer generalized velocity 
$(\bm U, \bm \Omega)$ as a function of the swimmer configuration and tail
spinning velocity $\Omega_t$.
Once $(\bm U, \bm \Omega)$ are known, the standard rigid body kinematic equations
can be solved for the swimmer head.  
The swimming problem is solved by decoupling the $(\bm U, \bm \Omega)$ 
into two parts where 
the active part $(\bm U_a, \bm \Omega_a)$ corresponds to the movement of the
microswimmer in a bulk fluid at rest 
while the passive part $(\bm U_p, \bm \Omega_p)$ corresponds to a passive swimmer ($\Omega_t=0$)
immersed in the external flow field $\bm u_b$. 
Thanks to the linearity, the active part can be expressed as
$\bm U_a = \Omega_t \bm R \tilde{\bm U}_{a}$ ,
$\bm \Omega_a = \Omega_t \bm R \tilde{\bm \Omega}_{a}$
where $\bm R$ is the rotation matrix that transforms the expression of a vector
in the body reference frame into its expression in the 
global reference frame and 
$(\tilde{\bm U}_{a}, \tilde{\bm \Omega}_{a})$ are the velocities for 
a microswimmer swimming with $\omega_{t} = 1$ in a configuration where 
body and global frame coincides.
Concerning the passive part, instead, 
we exploit the local decomposition of 
$\bm u_b$ in three components, 
a rigid translation at $\bm x_c$, 
a rigid rotation $\bm \omega_p^S$ (associated to 
antisymmetric part of the velocity gradient $S_{ij}$) 
and a deviatoric part 
(symmetric part of the velocity gradient $E_{ij}$). 
The deviatoric part $E_{ij}$ can be further decomposed into five components. 
For each of them, we can solve a swimming problem 
and get the contributions to the swimmer translational $\bm U$ 
and angular $\bm \Omega$ velocities.
Combining all those contributions, the microswimmer velocity in an external flow is
obtained as
\begin{align}
  \bm U      & = 
        \Omega_t \bm R  \tilde{\bm U}_{a} + \bm U_p^b + 
        \bm R \sum_{k=1}^5 \tilde \beta_k \tilde{\bm U}_k^{E}      \, , \label{eq_base_u}        \\
  \bm \Omega & = 
        \Omega_t \bm R  \tilde{\bm \Omega}_{a} + \bm \Omega_p^S + 
        \bm R \sum_{k=1}^5 \tilde \beta_k \tilde{\bm \Omega}_k^{E} \, . \label{eq_base_w}
\end{align}
where the first term on the right hand side is the active contribution, 
the second term is the uniform translation $\bm U_p^b$ and rotation 
$\bm \Omega_p^S$ due to the external flow and the last terms
are the five contribution due to the deviatoric part of
the velocity gradient. 
The weights $\tilde \beta_k$ depends only on $\bm u_b$ and
on the swimmer orientations, details are reported in the appendix~\ref{sec_apx_method}.
\tb{It is worth noting that, as a first approximation, 
a more appropriate model for the head-tail coupling is to fix 
the exchanged torque~\cite{berg1993torque,xing2006torque}.
However, in our case, the head-tail coupling enters only in the 
active part of~Eqs.\eqref{eq_base_u}-\eqref{eq_base_w}. 
Since the active part corresponds to the movement of the
microswimmer in a bulk fluid at rest, 
the tail spin $\Omega_t$ is proportional to the motor torque
and, hence, considering a fixed spin or a fixed torque only  amounts
to a linear rescaling with no effect on the observed phenomenology. 
}

The main advantage of the proposed method is that
only six solutions of the swimming problem
are needed;
one for $(\tilde{\bm U}_{a}, \tilde{\bm \Omega}_{a})$ and five 
for $(\tilde{\bm u}_k^{E}, \tilde{\bm \Omega}_k^{E})$.
These swimming problems can be solved with any Stokes solver. Here we use
the method of fundamental solution (MFS)~\cite{young2006method} 
that is summarized in the appendix~\ref{sec_apx_method}. 
Once these solutions are known, one can integrate the rigid body kinematics to
get the swimmer trajectory.
Here, this integration step is performed using a quaternion formulation 
and a $4^{th}$ order Runge-Kutta method.

%%%%%%%%%%%%%%%%%%%%%%%%%%%%%%%%%%%%%%%%%%%%%%%%%%%%%%%%%%%%%%%%%%%
\section{Results}

\tb{In this study, the microswimmer is immersed in an unbounded shear flow
\begin{align}
  \bm u_b &= (x_3 \tau_s, 0, 0)  \, .
\end{align}
Without loss of generality, we select as 
time unit $1/\tau_s$ and
as length unit of length $r_{h1}$ the larger axis of the ellipse.
Due to the linearity of the problem, the spin $\Omega_t$ 
is the only crucial parameter for given microswimmer.} 
For both the circular (model I) and the elliptical (model II) helical tail microswimmers,
we studied the motions at different tail spinning velocity $\Omega_t$.
For each $\Omega_t$, we simulated  from $10^3$ to $10^4$ trajectories starting from 
different initial conditions with random orientation.
The center of the head is initially placed in the origin at
$t=0$.
In all the cases, after a transient, 
the swimmer orientation converges to periodic trajectories. 
Concerning the swimmer translation, different scenarios are possible
depending on the swimmer tail geometry, 
\tb{its spinning velocity $\Omega_t$}
and
its initial condition. 
A summary of the different possibilities is reported in
figure~\ref{fig_phase_map} and discussed in the following sections.

\subsection{Circular helix}
For the circular helix swimmer, in the passive case (tail spinning velocity $\Omega_t = 0$)
after a transient, the swimmer is always 
oriented along $-\textbf{X}_2$, i.e. normally to the shear plane $\textbf{X}_1\textbf{X}_3$,
and it moves along $\textbf{X}_2$, i.e. $U_2 > 0$. 
In the Figure~\ref{fig_phase_map}a, those information are condensed  
in panel a) where the fraction of the trajectories that result
in final drift $U_2 > 0$ 
can be read on the left axes while the colored bars indicate the actual value
of $U_2$. For instance, the orange bar 
\tb{at $\Omega_t = 0$}
means that all the $10^3$
initial conditions result in a slightly positive terminal velocity $U_2 \cong  1.01 \times 10^{-4}$
while the blue bar 
\tb{at $\Omega_t = 0.015$}
indicates that almost all 
the swimmers reach a final velocity $U_2 \cong -3.00 \times 10^{-6}$.
Fig.~\ref{fig_phase_map}b, instead, reports the orientation $\eta$ averaged on a period.
For the pure passive case, $\Omega_t = 0$, we always get $\langle \eta \rangle \cong \pi$,
i.e. the swimmer is oriented perpendicularly to the shear plane.
This passive swimmer regime is indicated as $\mathrm{I_1}$ and a sketch
of its periodic orbits is reported in Fig.~\ref{fig_phase_map}e and in Video 
SM1~\footnote{See Supplemental Material at [URL will be inserted by publisher] for
movies showing these trajectories.}. 
This result is in agreement with the shear-induced 
separation of pure circular helix discussed in~\cite{fu2012bacterial} 
where it was shown that microswimmers point perpendicularly to the
shear plane in the direction here indicated as $-X_2$.
A similar behavior is also observed for low spinning velocity, 
\tb{$\Omega_t < 0.015$.}

A further increase of the tail spinning results in a first change in 
the dynamics. The average orientation of the swimmer is the same, 
$\langle \eta \rangle \cong \pi$,
but now the drift velocity is positive, $U_2>0$, $\mathrm{I_{2}}$ Fig.~\ref{fig_phase_map}f.
This is expected, indeed, as $\Omega_t$ increases, the swimmer propulsion
becomes more relevant until, finally, it dominates over the passive drift 
induced by the shear.
Interestingly, in some intervals 
of the spinning speed, an additional kinematics appears, 
$\mathrm{I_{3}}$ Fig.~\ref{fig_phase_map}g.
The swimmer undergoes to a Jeffery-like motion  
with $\langle \eta \rangle \in [0.8, 0.9] \pi$.
This motion is characterized by a slightly smaller value of the average velocity 
$U_2$.
Depending on the initial condition, 
some trajectories converge to a motion of the $\mathrm{I_{2}}$ kind and others to
$\mathrm{I_{3}}$.
Overall, those data indicate that 
the shear always orients the swimmer along $-\textbf{X}_2$.
For small tail spinning (passive case) the shear dominates the dynamics and the 
swimmer moves in the $\textbf{X}_2$ direction while,
for large tail spinning (active case), the self-propulsion dominates
and the swimmer moves in the $-\textbf{X}_2$ direction. 
%%%

\subsection{Elliptical helix}
A much richer scenario occurs for swimmers with an elliptical helix tail, model II, 
Fig.~\ref{fig_phase_map}c.
In the passive case, we observed three main different  
periodic trajectories. 
The overall drift is positive $U_2 > 0$ in this region, as for model I. 
The average orientation $\langle \eta/\pi \rangle$, however, is significantly different. 
The first kind of trajectory $\mathrm{II_{1}}$ orientates along $-\textbf{X}_2$ as for model I. 
The other two kinds of trajectories, $\mathrm{II_{2}}$ and $\mathrm{II_{3}}$, present 
\tb{Jeffery-like tumbling} behaviors that differ from $\mathrm{II_{1}}$, 
see Fig~\ref{fig_phase_map}h-j and Video SM2, SM3 and SM4~\cite{Note1}. 
In particular, for $\mathrm{II_{3}}$ we observe that the tumbling occours almost in the shear plane.
Similar to the shear induced separation of pure helix~\cite{fu2009separation}, 
such kind of tumbling (Jeffery-like) motion on the shear plane is
associated to a lateral velocity ($U_2$) of the microswimmer that, 
in our case, it is larger than the one corresponding to $\mathrm{II_{1}}$, see~\figref{fig_phase_map}c. 
\tb{No simple rules are found to associate the final 
microswimmer trajectory 
to its initial orientation, 
see Supplementary Section S1
~\footnote{See Supplemental Material at [URL will be inserted by publisher] for
a figure representing the dependence of the final stable trajectory 
on initial swimmer orientation.
} 
 where
examples of the time evolution of the orientation $\bf p$ are 
reported together with a diagram representing the domains in the orientation space
that led to $\mathrm{II_{1}}$, $\mathrm{II_{2}}$, $\mathrm{II_{3}}$ trajectories. 
 }

As spinning speed $\Omega_t$ increases, the system undergoes a gradual regularization. 
\tb{We still observe three different kinds of trajectory
but the values of the average orientation 
$\langle \eta/\pi \rangle$ of $\mathrm{II_1}$, $\mathrm{II_2}$ and $\mathrm{II_3}$ get closer,
until they merge. 
In this intermediate regime, 
trajectory $\mathrm{II_{1}}$ switch from positive to negative $U_2$
and, for this reason, we renamed it as $\mathrm{II_{4}}$,~\figref{fig_phase_map}l.   
 }
Further increases in $\Omega_t$ brings the system to a fully active regime
where only $\mathrm{II_{4}}$ trajectory is observed: 
the swimmer is oriented along $-\textbf{X}_2$ with $U_2 < 0$.
This regime is analogous to the active regime for circular helical tail, 
$\mathrm{I_{2}}$ trajectory.

%%%%%%%%%%%%%%%%%%%%%%%%%%%%%%%%%%%%%%%%%%%%%%%%%%%%%%%%%%%%%%%%%%%
\begin{figure}[]
  \includegraphics[width=\columnwidth]{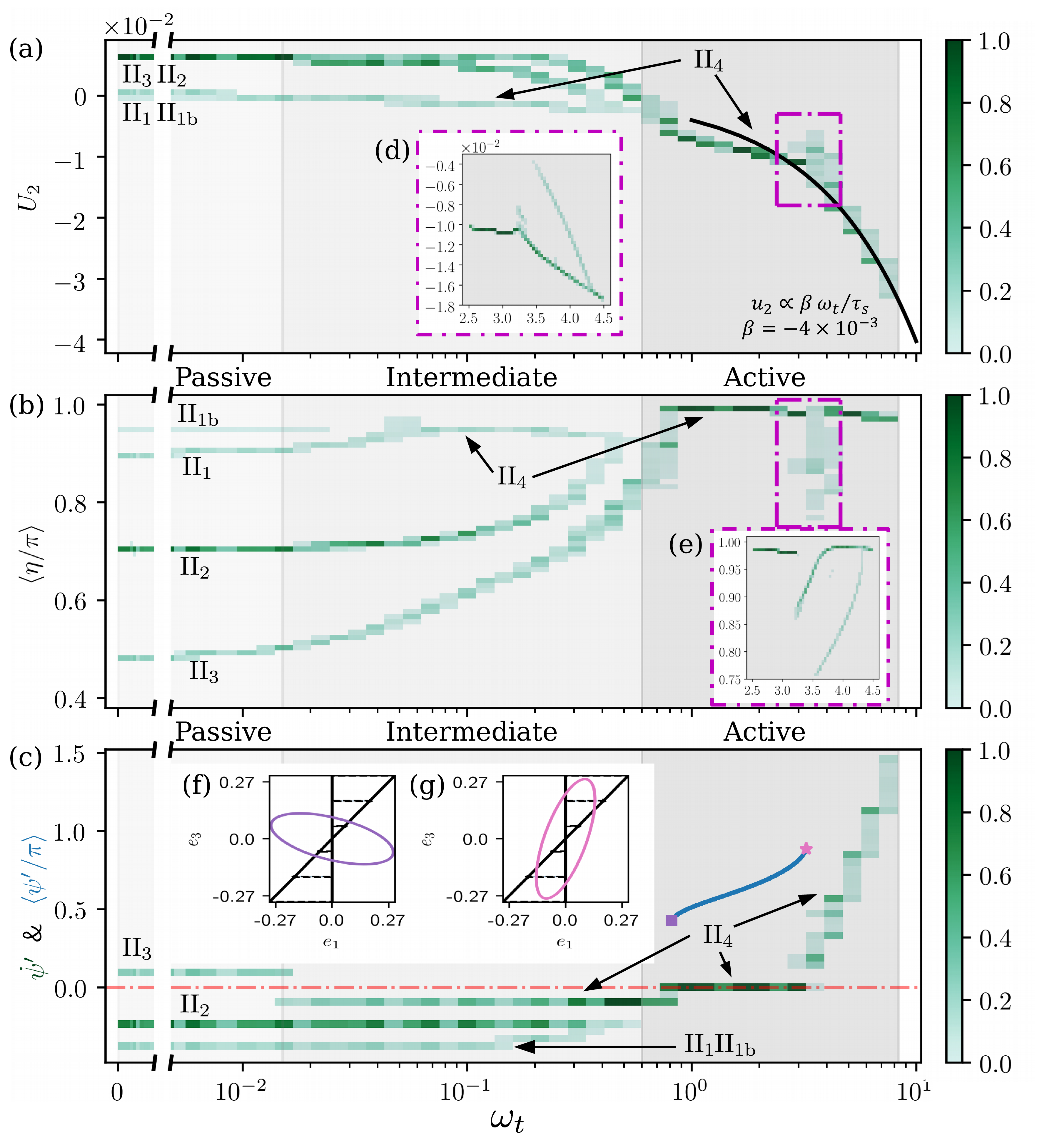}
  \caption{Elliptical helix tail microswimmer. 
Lateral velocity $U_2$ (a), 
normalized average angle $\langle \eta / \pi \rangle$ (b) 
and the absolute tail spin $\dot \psi' = \dot \psi + \dot \psi_t$ (c) 
\tb{as functions of $\Omega_t$.} 
Results refer to $450$ trajectories for each $\Omega_t$.
The green color scale indicates the probability that 
one initial condition converges to the corresponding
value on the vertical axis, for instance, in the passive case, almost $~75\%$ of
the swimmers converge to the trajectory $\mathrm{II}_2$ that 
corresponds to $\langle \eta \rangle \cong 0.7 \pi$ (dark green).
Black solid line in panel (a) is the velocity of the 
same microswimmer in a fluid at rest. 
Panels (d) and (e) report a detailed view of the regions enclosed by 
the violet dotted boxes.
The {\sl freezing} tail phenomenon discussed in the text is sketched
in panels (f) and (g) while the corresponding average  $\psi'$ is reported as a solid blue line
\tb{between $0.8 <\Omega_t < 3.2$ in panel (c).}
}
\label{fig_eval_modII}
\end{figure}
%%%%%%%%%%%%%%%%%%%%%%%%%%%%%%%%%%%%%%%%%%%%%%%%%%%%%%%%%%%%%%%%%%%

To better characterize the elliptical helical tail microswimmer,
we performed additional simulations that allowed us to
observe further details of the swimmer motion.
Results are reported in~\figref{fig_eval_modII}a for the lateral velocity $U_2$
and in~\figref{fig_eval_modII}b for the normalized average angle $\langle \eta / \pi \rangle$.
\tb{For each $\Omega_t$}
we performed 450 simulations with random initial orientation.
The green color scale corresponds to the probability that the swimmers reach 
a steady state with the corresponding value of $U_2$ and $\langle \eta / \pi \rangle$. 
For instance, at
\tb{low $\Omega_t$}
 (passive regime), four kinds of stable trajectories 
exist. Three of them, $\mathrm{II}_1, \mathrm{II}_2, \mathrm{II}_3$, 
were already discussed in~\figref{fig_phase_map}.
The last one, indicated as $\mathrm{II}_{1b}$, 
corresponding to $\langle \eta \rangle \sim 0.95 \pi$, 
is quite rare (light green in~\figref{fig_eval_modII}b) and very similar to $\mathrm{II}_1$. 
As already discussed in Fig~\ref{fig_phase_map}, the trajectories oriented perpendicularly to the shear plane 
($\mathrm{II}_1$ and $\mathrm{II}_{1b}$ for which $\langle \eta \rangle \in [0.9, 1] \pi$)
 have almost no lateral motion ($U_2 \cong 0$). 
In contrast, the other two kinds of trajectories, 
characterized by \tb{Jeffery-like tumbling} close to the shear plane ($\mathrm{II}_2$ and $\mathrm{II}_3$), 
show a significant lateral motion, $U_2 > 0$,
see also~\figref{fig_phase_map}i and~\figref{fig_phase_map}j.
Moreover, \figref{fig_eval_modII} also better evidences how, 
through increasing of tail spin $\Omega_t$, the tumbling trajectories $\mathrm{II}_2$ 
and $\mathrm{II}_3$ 
progressively converge towards the $-X_2$ axes as apparent  
from the increase of $\langle \eta / \pi \rangle$.
Finally, in the active regime, all the trajectories merge into 
a single kind where the swimmer is oriented normal to the shear plane
$\langle \eta \rangle \cong \pi$. 
\subsection{Freezing spin}
Nevertheless, some islands of complexity persist in this active region.
For instance, the microswimmer is {\sl frozen} by the shear flow for 
\tb{spinning $0.8 <\Omega_t < 3.2$.} 
The tail of the microswimmer, when seen from the global reference frame, 
does not spin along the swimmer axis $\bm p$. 
This is apparent in~\figref{fig_eval_modII}c
where the time derivative of the angle $\psi'$ is reported.
In essence, the tail rotates with respect to the head 
($\dot \psi_t = \Omega_t$ is imposed in our model) but
the rotation of the head with respect to the global reference frame
exactly counter balances the spinning 
($\psi'=\psi + \psi_t$, 
 $\dot \psi'= \dot \psi + \dot \psi_t$, hence, 
$\dot \psi'= 0 \Rightarrow \dot \psi =  -\dot \psi_t$), see Supporting video~SM5~\cite{Note1}.
This is a peculiar behavior that occurs only for the elliptical helical tail and not 
for the circular one and it represents a further indication that
slight changes in the swimmer geometry may lead to new phenomena.
In fact, the tail of the microswimmer experiences a propulsion torque due to propulsion 
as well as a shear torque due to local velocity gradient. 
The balance between the two torques on the tail leads to the {\sl freezing}.
For the lowest spinning velocity for which the {\sl freezing} occurs, 
\tb{i.e. $\Omega_t = 0.81$, }
the propulsion torque is small. 
Thus, the mayor axis of the tail section is almost
parallel to the shear velocity direction and, consequently, 
the torque induced by the shear on the tail  is small, as in~\figref{fig_eval_modII}f. 
As the tail spinning $\Omega_t$ increases, the propulsion torque increases and the new balance 
is found for larger values of $\psi'$. 
The maximum shear torque is achieved when the mayor axis of the tail section is vertical and, 
indeed, the last value 
\tb{of $\Omega_t = 3.24$}
for which this tails freezing occur corresponds to $\psi' \approx \pi$, see~\figref{fig_eval_modII}g. 

Another unexpected behavior occurs for 
\tb{$\Omega_t \in (3, 4.5)$} 
where 
we observe that, again, the swimmer may converge 
towards multiple different trajectories, see~\figref{fig_eval_modII}d and~\figref{fig_eval_modII}e.
\tb{All these trajectories have a negative $U_2$ and their oscillation 
around $\textbf{X}_2$ axis is limited, $\langle \eta /\pi \rangle > 0.7$.
For these reasons they can be overall classified as $\mathrm{II_4}$.}
Only after this last region of complexity,
the motion gets finally regularized. In this fully active regime,
the final swimmer speed is linear in the tail spinning, 
\tb{$U_2 = \beta \Omega_t$, }
with $\beta = -4.02 \times 10^{-3}$. 
This is expected, indeed, when the tail spin is large, the 
final swimmer speed is dominated by the propulsion.
Indeed, the value of $\beta$ we observed is the same as we got in 
a simulation of the active swimmer moving in a fluid at rest 
represented as a black solid line in~\figref{fig_eval_modII}a.
In essence, in the active regime, the shear selects the swimmer orientation, 
and the final speed is controlled by the tail spin. 
In the active regime, the swimmer dynamics is predictable and controllable:
any initial condition results in the same final trajectory.

%%%%%%%%%%%%%%%%%%%%%%%%%%%%%%%%%%%%%%%%%%%%%%%%%%%%%%%%%%%%%%%%%%%%%%%%%%%%%%
\section{Conclusion}

In this manuscript, we proposed an efficient computational method for
the analysis of microswimmer motion in external flows.
We applied our method for the analysis of  microswimmers whose propulsion
is due to the spinning of a flagellum ({\sl E.coli}-like swimmers).
Once the swimmer geometry is selected, the entire range of spinning speed of the tail 
can be explored by solving only six swimming problems.
This allowed us to simulate thousands of different trajectories.
We compared the motions of two different swimmers, one
carrying a circular helical tail, 
i.e. a helix that lies on a circular cylinder, that is the typical geometry 
studied in previous theoretical and computational works,
and another one carrying an elliptical helical tail. 
The alteration of the tail shape from circular helix to elliptical helix 
gives rise to a much richer scenario where different tumbling 
(Jeffery-like) trajectories can be observed under the same external flow condition
and for the same tail spinning speed.
As the propulsion torque increases, a progressive regularization
of the motion is observed until, in the propulsion dominated regime,
the swimmers converge to the same final trajectory for all the initial configurations.
These results may have some implications on the biology of microorganisms that exploit this
propulsion mechanism.
Indeed, the complex \tb{Jeffery-like tumbling} we observed in the shear dominated regime may provide an alternative
way to increase the capability of a microswimmer to explore the space 
that may cooperate with the well known {\sl run and tumble} motion~\cite{berg2000motile}. On the other hand, the high sensitivity to the shape of the tail implies that the microorganism must reach a larger spinning frequency in order to have a full control of its asymptotic swimming direction.
As a result, the presence of more that one steady state also has to be
carefully taken into account when designing artificial microswimmers
whose motion in external flows needs to be controlled.

%%%%%%%%%%%%%%%%%%%%%%%%%%%%%%%%%%%%%%%%%%%%%%%%%%%%%%%%%%%%%%%%%%%%%%%%%%%%%%

\begin{acknowledgements}
The authors would like to thank Prof. Yang Ding and Prof. Xinliang Xu
for useful discussion on the computational approach. 
This project was supported by the program of China Scholarships Council (No. 201804890022).
\end{acknowledgements}

%%%%%%%%%%%%%%%%%%%%%%%%%%%%%%%%%%%%%%%%%%%%%%%%%%%%%%%%%%%%%%%%%%%%%%%%%%%%%%
\appendix

\section{Details on the methods} 
\label{sec_apx_method}

In this appendix, 
we discuss the approach we employed for the solution of the 
swimming problem for an active microswimmer with a single intrinsic degree of 
freedom (DOF) swimming in an external flow. 
The DOF is the spin of the tail with respect to the microswimmer body. 
This model can be easily extended to multiple DOFs. 
Our method is a combination of known approaches for solution of the
Stokes equation that, for completeness, are reported in the following sections.
The crucial idea it to decompose the rate of strain 
in five base components. This allows to reduce the solution of
the swimming problem to six solutions of the Stokes equation,
one for the active propulsion 
and five for the passive one.
These swimming problems can be solved with any Stokes solver. Here we employed
the method of fundamental solution (MFS)~\cite{young2006method} 
\tb{
Before entering in the details of our formulation, we briefly 
mention some alternative approaches for the swimming problem.
}

\tb{
Modeling the motion of a microswimmer 
using multiple rigid bodies 
is a relatively common approach 
(see, e.g.~\cite{shum2010modelling,pimponi2016hydrodynamics}).
A key to calculate the trajectory of a microswimmer 
is to compute the generalized velocity $(\bm U, \bm \Omega)$,
that can be calculated solving the Stokes equations plus the force- and torque-free 
conditions~\cite{elgeti2015physics}.
The boundary element method 
is commonly used for Stokes 
equations~\cite{shum2010modelling, liu2014propulsion},
although the solving method can be replaced by other formulations, such as
the method of 
regularized Stokeslets~\cite{rorai2019limitations,cortez2005method, zhang2020active}, 
the boundary integral method~\cite{klaseboer2012non}, 
and the spectral boundary element method~\cite{muldowney1995spectral}. 
Since usually it is
computationally expensive to calculate the generalized velocity 
$(\bm U, \bm \Omega)$ directly using full solution of the Stokes equation,
several approximate theories were developed for rigid body motion
in Stokes flows.
Following Marcos et. al. work~\cite{fu2009separation, fu2012bacterial},
Mathijssen et. al.~\cite{mathijssen2019oscillatory}
developed an approximate 
formulation of an ideal chiral object 
using the resistive force theory that allowed
to study bacteria rheotaxis
close to a surface.
Another alternative approach
is to calculate the 
generalized mobility matrix
of the system~\cite{kim1991microhydrodynamics}.
For a three sphere swimmer model~\cite{najafi2004simple},
a quadrupole order accurate multipole expansion
was recently employed to study the swimmer kinematics
close to a wall under a shear flow~\cite{daddi2020tuning}.
The possibility to extend this promising approach to more complex
swimmer geometries is, however, an open issue.
}

%%%%%%%%%%%%%%%%%%%%%%%%%%%%%%%%%%%%%%%%%%%%%%%%%%%%%%%%%%%%%%
\subsection{Fundamental solution of Stokes equation}
%%%%%%%%%%%%%%%%%%%%%%%%%%%%%%%%%%%%%%%%%%%%%%%%%%%%%%%%%%%%%%

%%%%%%%%%%%%%%%%%%%%%%%%%%%%%%%%%%%%%%%%%%%%%%%%%%%%%%%%%%%%%%%%%%%
\begin{figure}[]
  \includegraphics[width=\columnwidth]{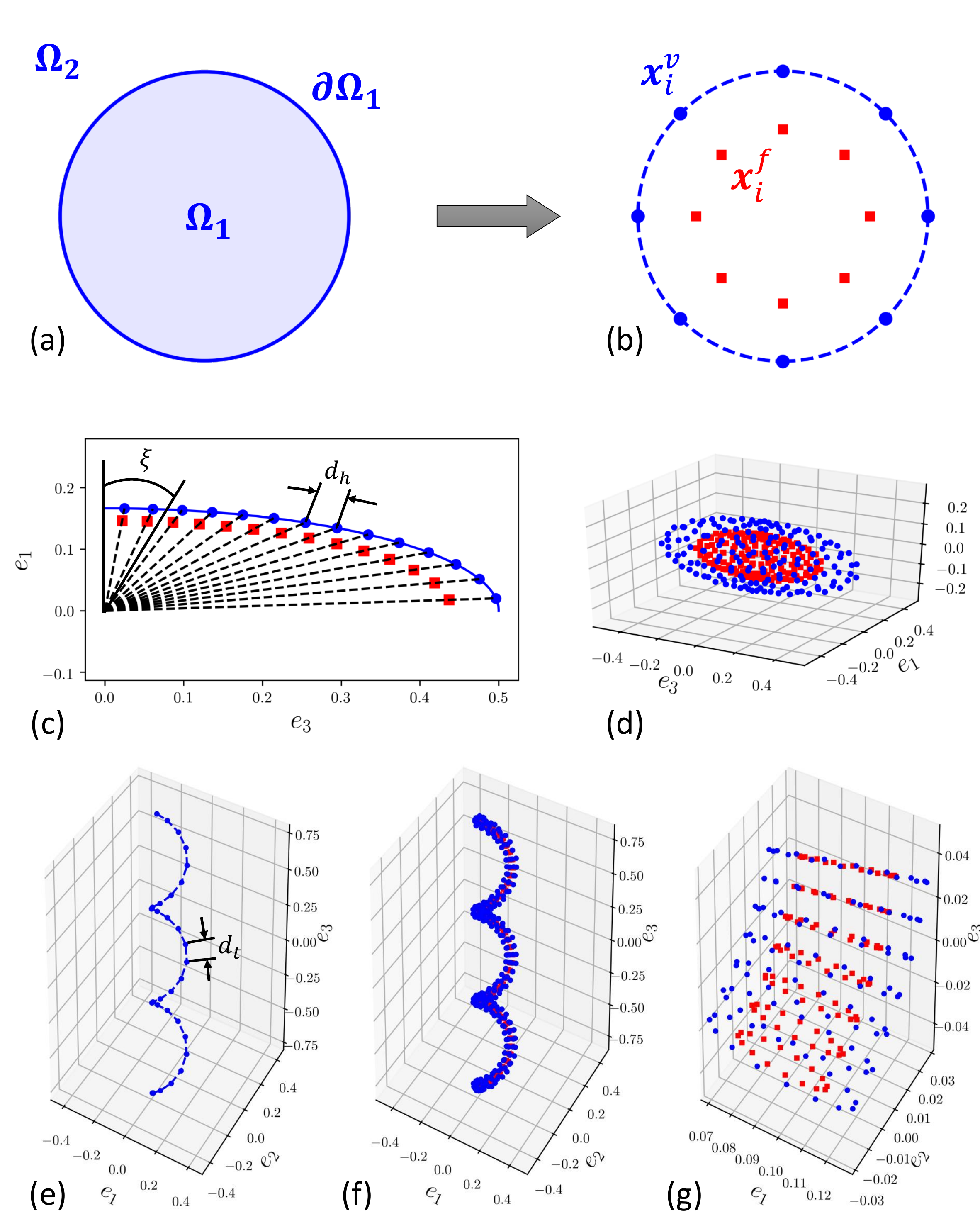}
  \caption{
    Sketch of the method of fundamental solutions. 
    (a) A solid body (domain $\Omega_1$) moves in a bulk fluid (domain $\Omega_2$). 
    The solid blue line $\partial \Omega_1$ is the boundary of the solid body. 
    Boundary points $\bm x_i^v$ \tb{(blue circles)} are selected on $\partial \Omega_1$ while 
    source points, $\bm x_i^f$ \tb{(red squares)} are placed inside the solid body. 
    Panel (c) shows the discretization used for a quarter of an ellipse while
    the swimmer head is in the panel (d).
    Panel (e) reports the tail centerline while panel (f) refers to the 
    discretization of the  swimmer tail.
    Each section of the tail is modelled as a circle where, again, \tb{red squares 
    correspond to force sources and blue circles to the boundary}. In panel (g) a 
    short section of the swimmer tail is shown.
  } 
  \label{fig_apx_MFS}
\end{figure}
%%%%%%%%%%%%%%%%%%%%%%%%%%%%%%%%%%%%%%%%%%%%%%%%%%%%%%%%%%%%%%%%%%%

Here, we briefly summarize the method of fundamental solution (MFS)~\cite{young2006method}.
In the creeping flow limit, the governing equation for the 
fluid velocity $\bm{u}$ and pressure $p$ 
due to a point force singularity 
of strength $\bm{f}$ applied to the point $\bm x_f$
is the Stokes equations 
\begin{align}
    \nabla\cdot\bm{u} &= 0 \, , 
    \label{eq_apx_01}\\
    \mu\nabla^2\bm{u} &= \nabla p-\bm{f}(\bm x_f)\delta(\bm{x} -\bm x_f) \, , 
    \label{eq_apx_02}
\end{align}
where $\mu$ the fluid viscosity, 
and $\delta$ is the Dirac delta function.
The solution  
of~\eqref{eq_apx_01}-\eqref{eq_apx_02} (also knows as Stokeslet) reads
\begin{equation}
    \bm u(\bm x) = \bm{S}(\bm x_f, \bm x) \bm f(\bm x_f) \, , 
\end{equation}
with 
\begin{equation}
    \bm{S}(\bm x_f, \bm x) 
    = 
    \frac{1}{8\pi\mu}
    \left( 
       \frac{\textbf{I}}{r}+\frac{(\bm x_f-\bm x)(\bm x_f-\bm x)}{r^3}
    \right) \, ,
\end{equation}
where $\textbf{I}$ is the unit matrix, 
and $r = \norm{\bm x_f -\bm x}$.
The tensor $\bm{S}(\bm x_f, \bm x)$ is commonly indicated as Oseen tensor.

The MFS~\cite{young2006method} was already successful employed 
in microfluidics, see e.g.~\cite{aboelkassem2013stokeslets,lockerby2016fundamental}. 
In brief, as is shown in \figref{fig_apx_MFS} (a, b), for problems where the velocity is 
assigned on the boundary of a solid domain $\Omega_1$ and
the velocity field needs to determined in the external domain 
$\Omega_2$, 
the key of the MFS is to find an approximation field 
$\bm{u'}$ that is defined in the domain $\Omega_1 \cup \Omega_2$ and
that fulfills the boundary condition at the frontier of $\Omega_1$.  
The fluid velocity field $\bm{u'}$ is a smooth field that is 
defined in both the domains $\Omega_1$ and $\Omega_2$.
A set of $n$ boundary points $\bm x_i^v$ located at the boundary $\partial \Omega_1$ 
are selected. For each one of them, we know its corresponding velocity 
$\bm{u}(\bm x_i^v)$ from boundary conditions.
A set of $n$ point forces are placed inside the domain $\Omega_1$ close to the boundary points,
the location of the point forces being indicated as $\bm x_i^f$. 
Hence, the velocity $\bm{u}(\bm x_i^v)$ can be expressed as
%%%
\begin{equation}
  \begin{bmatrix}
      \bm{u}(\bm x_1^v)\\ 
      \ldots\\ 
      \bm{u}(\bm x_n^v)
  \end{bmatrix}=
  \begin{bmatrix}
      \bm{S}(\bm x_1^f, \bm x_1^v) & \ldots & \bm{S}(\bm x_n^f, \bm x_n^v) \\
      & \ddots & \\
      \bm{S}(\bm x_1^f, \bm x_1^v) & \ldots & \bm{S}(\bm x_n^f, \bm x_n^v) \\
  \end{bmatrix}
  \begin{bmatrix}
      \bm{f}(\bm x_1^f)\\ 
      \ldots\\ 
      \bm{f}(\bm x_n^f)
  \end{bmatrix} \, . 
  \label{eq_UMF}
\end{equation}
%%%
This system has $3n$ unknowns and $3n$ equations.
Once~\eqref{eq_UMF} is solved for $\bm f$, the approximated velocity $\bm{u'}$ in a
generic point $\bm x$ of the domain $\Omega_1 \cup \Omega_2$ can be calculated as
\begin{equation}
  \bm{u'}(\bm{x}) 
  =
  \sum_{i=1}^n \bm{S}(\bm x_i^f, \bm{x}) \bm{f} (\bm x_i^f) \, .
\label{eq_vel}
\end{equation}
In the following, to simplify the notation, we will use the same symbol $\bm{u}$ for the 
approximated velocity and the true solutions of the Stokes problem.

%%%%%%%%%%%%%%%%%%%%%%%%%%%%%%%%%%%%%%%%%%%%%%%%%%%%%%%%%%%%%%
\subsection{The discretization of the microswimmer}
%%%%%%%%%%%%%%%%%%%%%%%%%%%%%%%%%%%%%%%%%%%%%%%%%%%%%%%%%%%%%%
A technical issue in MFS concerns the location 
$\bm {x}_i^f$ of the point sources.
Our swimmer is composed by a spheroidal head and a helical tail. 
Concerning the head, we first placed the boundary point 
on a 2D ellipse with semi-axes $r_{h1}$ and $r_{h2}$ 
lying on the $\textbf{e}_3 \textbf{e}_1$ plane, approximately 
\tb{at the same distance $d_h$,
\footnote{Considering one-quarter 
of an ellipse, the arc length is a monotone increasing function of 
the eccentric angle $\xi$ that has no explicit expression.
Therefore, we first fit this function using a quadratic polynomial, 
then determine a set of $\xi_i, i \in (1, n_h)$ that keeps 
the distance between two adjacent points approximately equal.
Finally we calculate the location of the points on the 
$\textbf{e}_3 \textbf{e}_1$ plane, see~\cite{Andy2020}}
\cite{Andy2020},}
 see~\figref{fig_apx_MFS}(c).
The ellipsoid is a body of revolution. 
Hence, we rotated each point around the major axis $\textbf{e}_3$ 
of the ellipsoid obtaining a circle perpendicular to $\textbf{e}_1$. This circle 
is divided into boundary points with equal distance $d_h$.
In this study, we select $d_h / r_{h1} \cong 0.047$ for a total of $1653$ 
boundary points lying on the swimmer head and indicated as $\bm x_i^{vh}$.  

For each boundary point $\bm x_i^{vh}$, a point force is located inside the ellipsoid on the 
lines that connect $\bm x_i^{vh}$ with the ellipsoid center $\bm x_c$.
The distance $r_i^{fh}$ between $\bm x_c$ and $\bm x_i^{fh}$ 
is given by
\begin{equation}
  r_i^{fh} = \delta_h r_i^{vh} \quad  , \quad  \, 
  \delta_h = 1 + \dfrac{ 2 \epsilon _h \langle d_h \rangle}{(r_{h1} + r_{h2})} \, , 
\end{equation}
where $r_i^{vh}$ is the distances between the ellipse center 
$\bm x_i^{vh}$, $\langle d_h \rangle$ is the average 
distance of the neighbor boundary points and 
$\epsilon_h$ is a control parameter. 
In this study, we used $\epsilon_h = -1$. 
We also verified that results does not change for $\epsilon _h \in (-0.5, -1)$. 
\figref{fig_apx_MFS} (d) shows a example of the ellipsoid after discretization. 

Concerning the tail, we first defined its centerline in
a reference system with origin in the swimmer head center $\bm x_c$ as
\begin{equation}
  {\bm r_t(s)} = (r_{t1} \cos(2 \pi s), r_{t2} \sin(2 \pi s), \lambda s - \delta_{bt}) \, ,
\end{equation}
where $s \in [-n/2, n/2]$ 
with $n$ the number of periods of the tail, 
$\lambda$ is the pitch of the helix
$\delta_{bt}$ is the distance from $\bm x_c$ to the tail center, 
here set to \tb{$\delta_{bt} = r_{h1} + n \lambda/2 + r_{h1}/2$}, 
and $r_{t1}$ and $r_{t2}$ are the radius of the elliptical cylinder
on which the helix lies.
\tb{We also performed a set of simulations analogous to the 
ones discussed in Fig~\ref{fig_eval_modII} but with 
$\delta_{bt} = r_{h1} + n \lambda/2 + r_{h1}/5$.
Beside minor quantitative differences, the results 
fairly agree with the one discussed in the manuscript. 
}
We discretize $s$ into $m+1$ values $s_i = -n/2 + i n/m , i \in (0, m)$, as shown in~\figref{fig_apx_MFS}(e). 
Then, for each of them, we put a circle of radius $\rho_t$ 
perpendicular to the centerline of the helix. 
This circle is divided into boundary points with equal distance $d_t$
The associated point force are placed on the concentric circle that 
perpendicular to the helix centerline, as is shown in~\figref{fig_apx_MFS}(b). 
The radius of this concentric circle is $\rho_t - \epsilon_t d_t$.
with $\epsilon_t = -1$. 
In this study, we select $d_t = \sqrt{\lambda^2 + C_{elp}^2} n / m \cong 0.019$, 
where $C_{elp}$ indicates the perimeter of the ellipse with radius $r_{t1}$ and $r_{t2}$. 
The two ends of the helix are closed using semi-spheres. 
The generation method of the discretized semi-sphere is the same as one 
used for the ellipsoidal head of the microswimmer where, now, we used 
$r_{h1} = r_{h2} = \rho_t$ while $d_t$ is the distance among the boundary points
of the hemi-sphere. 

Setting as unit of length the larger axis of the ellipse, the circular 
helical tail microswimmers has the following geometrical parameters
$r_{h1} = 1/2$, $r_{h2} = 1 / 6$, $r_{t1} = 0.1$, $r_{t2} = 0.1$, $\rho_t = 0.03$,
$n = 3$, $\lambda = 2 / 3$. The number point forces is 
$1653$ for the head and $1534$ for the tail.
For the elliptical helical tail all the parameters are the same 
as for the circular tail swimmer with the exception of 
$r_{t1} = 0.3$. The number of point forces on elliptical helix tail is 2464.

%%%%%%%%%%%%%%%%%%%%%%%%%%%%%%%%%%%%%%%%%%%%%%%%%%%%%%%%%%%%%%
\subsection{Swimmer kinematics and boundary conditions}
%%%%%%%%%%%%%%%%%%%%%%%%%%%%%%%%%%%%%%%%%%%%%%%%%%%%%%%%%%%%%%

The microswimmer has seven degrees of freedom (DOFs), six 
DOFs represent the rigid motion of the head while the other the 
spinning of the tail.
Without loss of generality, for the translational DOFs we selected 
the center $\bm{x_c}$ of the ellipsoid that constitutes the swimmer head,
while for the orientational  DOFs, we selected the angles $\phi$, $\theta$ and
$\phi$ reported in~\figref{fig_setup}. 
The associated translational and rotational 
velocity are here indicated as
$\bm U$ and  $\bm \Omega$.
The tail rotates around the swimmer axis $\bm p \equiv \bm e_3$
at a spinning rate $\Omega_t$ with respect to the head. 
The no-slip boundary condition is applied on the surfaces 
of the head and the tail of the microswimmer, hence,
the fluid velocity at the swimmer boundary point is
\begin{align}
  \bm u(\bm x_i^{vh}) &= \bm U + \bm \Omega \times \bm r_i^{vh} \, , 
  \label{eq_apx_bc_head} \\
\end{align}
for the head boundary points $\bm x_i^{vh}$ and
\begin{align}
  \bm u_t(\bm x_i^{vt}) &= \bm U + (\bm \Omega + \Omega_t \bm p) \times \bm r_i^{vt} \, , 
  \label{eq_apx_bc_tail}
\end{align}
where in both equations $\bm r_i$ indicates the relative position of 
the boundary point with respect to the center of the swimmer head $\bm x_c$.
In our problem, 
the tail spin $\Omega_t$ is given and the other six DOFs are unknown. 
Thus, applying~\eqref{eq_apx_bc_head} and~\eqref{eq_apx_bc_tail} 
into~\eqref{eq_UMF}, 
we get a system of $3n$ variables in $3n+6$ unknowns. 
To complete this problem, we needed additional 
six equations that are the force- and torque- free 
conditions of the microswimmer 
\begin{align}
    \sum_{i=1}^{n_h} \bm f(\bm x_i^{fh}) + 
    \sum_{i=1}^{n_t} \bm f(\bm x_i^{ft}) &= 0\, , 
    \label{eq_force_free} \\
    \sum_{i=1}^{n_h} \bm r_i^{fh} \times \bm f(\bm x_i^{fh}) + 
    \sum_{i=1}^{n_t} \bm r_i^{ft} \times \bm f(\bm x_i^{ft})&= 0\, , 
    \label{eq_torque_free}
\end{align}
obtaining a system of $3n+6$ variables in $3n+6$ unknowns. 

The system was solved using the GMRES method~\cite{saad1986gmres} implemented 
in PETSc~\cite{balay1997efficient, petsc-web-page}. 
The solution provides the 
the rigid body translational $\bm U$ and rotational  $\bm \Omega$ 
velocities of the microswimmer head and the 
$3n$ components of the point force, from
which, using~\eqref{eq_vel} the entire velocity field can
be build.

Once the swimmer head generalized velocity $(\bm U , \bm \Omega)$ 
is obtained, the swimmer configuration is 
updated using the following kinematic equations
\begin{align}
  \dfrac{d \bm x_c}{dt} & = \bm U \, , 
  \label{eq_kinematic1}   \\
  \dfrac{d \bm e_i}{dt} & = \bm \Omega \times \bm e_i \quad , \quad i={1,2,3} \, , 
  \label{eq_kinematic2} \\
  \dfrac{d \psi_t}{dt}  &  = \Omega_t \, . 
  \label{eq_kinematic3}
\end{align}
As commonly did in microswimmer 
problems~\cite{shum2010modelling,pimponi2018flagellated}, 
in our code, we replaced~\eqref{eq_kinematic2} with
the quaternion formulation~\cite{graf2008quaternions,diebel2006representing}, 
to keep a higher numerical accuracy.
Eq.~\eqref{eq_kinematic1}-\eqref{eq_kinematic3} were solved
using a $4^{th}$ order Runge-Kutta 
method~\cite{bogacki1996efficient} implemented 
in PETSc~\cite{abhyankar2018petsc, petsc-web-page}. 

%%%%%%%%%%%%%%%%%%%%%%%%%%%%%%%%%%%%%%%%%%%%%%%%%%%%%%%%%%%%%%
\subsection{The method of base flow}
%%%%%%%%%%%%%%%%%%%%%%%%%%%%%%%%%%%%%%%%%%%%%%%%%%%%%%%%%%%%%%
In principle, the swimming problem presented in the 
previous section needs to be solved at any time step 
of the Runge-Kutta integrator used to update the swimmer configuration. 
This will require a large amount of computational resources.
Here we present an approach to largely speed up the simulation.
This approach is based on the decomposition 
of the swimmer motion into two parts, an active 
part and a passive part.
The idea of motion decomposition in the creep limit has a long history. 
For example, the motion of a particle in Stokes flow can be decoupled 
into the translation and the rotation parts~\cite{kim1991microhydrodynamics, happel2012low}. 
\tb{
Using this approach, Chwang and Wu~\cite{chwang1975hydromechanics} derived several 
exact solutions of the motion of a spheroid in a Stokes flow.
Subramanian and Koch extended their work and 
discussed the orientation of a passive spheroid 
in the simple shear 
flow~\cite{subramanian2006inertial, banerjee2020anisotropic}
and planar linear flow~\cite{marath2018inertial}.
Analytical solutions of the microswimmer motion 
with arbitrary geometry in the five basis flows, however, is difficult.
Hence, after decomposing the motion, we employed the numerical method
of the fundamental solution (described in the previous
section) to solve the Stokes problems. 
}

More specifically, we decouple the swimmer kinematics as it follows:
i) the active part $(\bm U_a, \bm \Omega_a )$
corresponding to the microswimmer self-propelling
in a bulk fluid at rest, and
ii) the passive part $(\bm U_p, \bm \Omega_p)$ corresponding
to a passive microswimmer (i.e. no tail spinning, $\Omega_t = 0$) in 
an external flow $\bm u_b$.
In formulae,
\begin{align}
  \bm U(\bm x_c, \bm \theta_c, \psi_t) 
      & = \bm U_a(\bm \theta_c, \psi_t) + \bm U_p(\bm x_c, \bm \theta_c, \psi_t) \, ,     \\
  \bm \Omega(\bm x_c, \bm \theta_c, \psi_t) 
      & = \bm \Omega_a(\bm \theta_c, \psi_t) + \bm \Omega_p(\bm x_c, \bm \theta_c, \psi_t) \, .
\end{align}
where we collectively indicated with $\bm \theta_c$ the three angles $\theta$, $\phi$ and $\psi$, 
see~\figref{fig_setup} defining the swimmer orientation.

%%%%%%%%%%%%%%%%%%%%%%%%%%%%%%%%%%%%%%%%%%%%%%%%%%%%%%%%%%%%%%%%
\begin{figure}[b]
  \centering
  \includegraphics[width=\columnwidth]{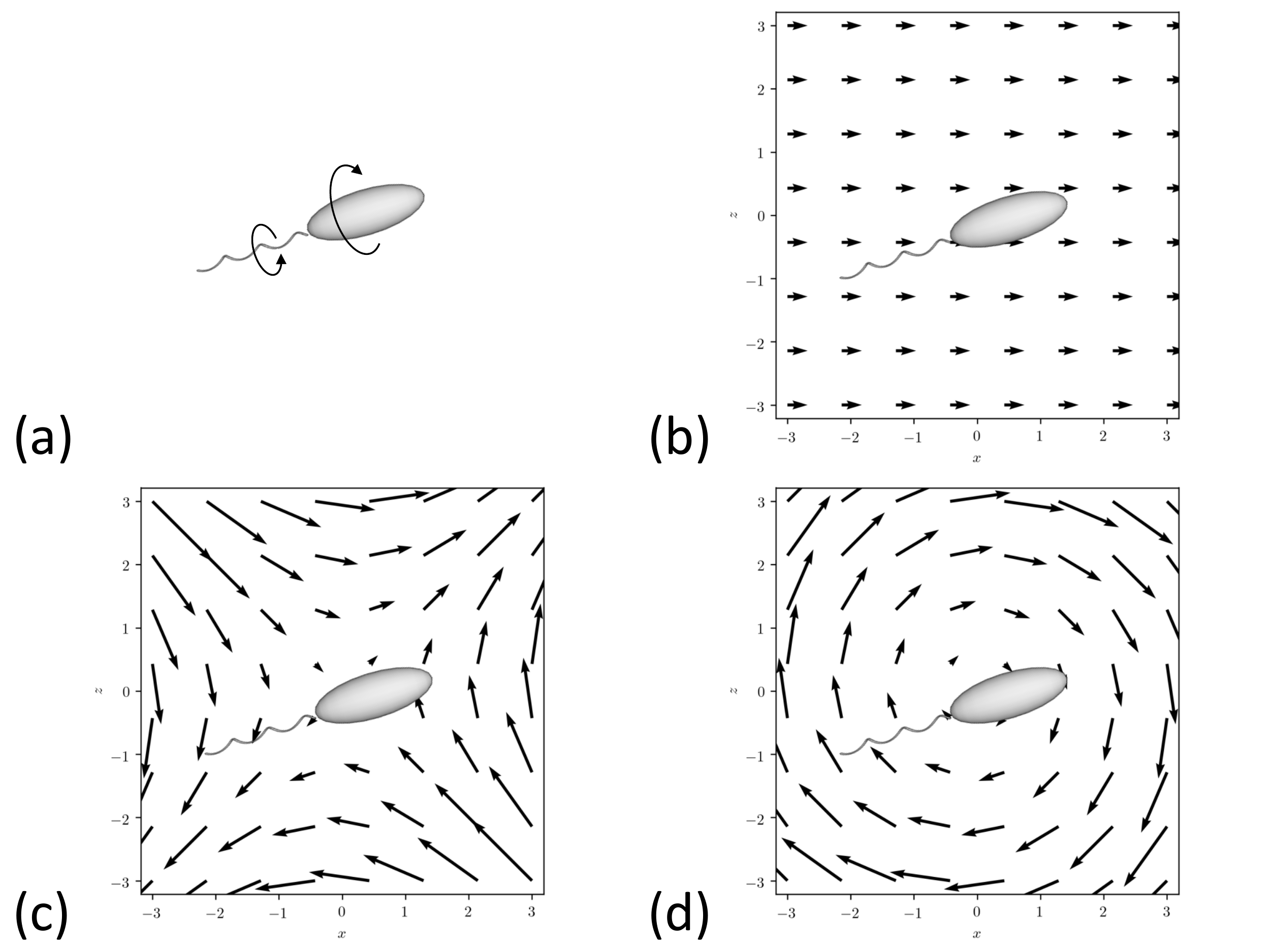}
  \caption{Sketch of the kinetic decoupling of the microswimmer in a external flow.
    (a) Active microswimmer motion $(\tilde{\bm U}_{a}, \tilde{\bm \Omega}_{a})$ in bulk fluid at rest.
    (b) Passive microswimmer translation $\bm u_b$ in the external flow.
    (c) Passive microswimmer motion $(\tilde{\bm U}^{E}, \tilde{\bm \Omega}^{E})$ in the symmetric (deviatoric) part of the external flow.
    (d) Passive microswimmer rotation $\bm \Omega_p^S$ in the antisymmetric part of the external flow.
  }
  \label{fig_apx_decoupling}
\end{figure}
%%%%%%%%%%%%%%%%%%%%%%%%%%%%%%%%%%%%%%%%%%%%%%%%%%%%%%%%%%%%%%%%

{\sl Active motion.} For the active part,
we first numerically calculated 
the unit-spin motion $(\tilde{\bm U}_{a}, \tilde{\bm \Omega}_{a})$ of 
a microswimmer swimming with $\omega_{t} = 1$ 
pointed toward the $\textbf{X}_3$ direction with 
$(\theta=0, \phi=0, \psi=0, \psi_t=0)$.
Thanks to the rotational symmetry of the ellipsoidal head, 
the last 2 rotational DOFs can be reduced to single DOF $\psi'=\psi+\psi_t$.
Indeed, if we take a given conformation on the swimmer
and we applied a rotation of the entire swimmer of an angle $\psi = \alpha$ and
then a rotation of the  tail with respect to the head of and angle $\psi_t = -\alpha$
the initial and the final conformations are the same.
Therefore, we can easily transform the motion $(\bm U_a, \bm \Omega_a)$ 
of an active swimmer whose tail spins at a rate $\Omega_t$ from the body coordinate frame 
$\textbf{O}'\textbf{e}_1\textbf{e}_2\textbf{e}_3$ 
to the global coordinate frame $\textbf{O}\textbf{X}_1\textbf{X}_2\textbf{X}_3$
\begin{align}
  \bm U_a (\bm \theta_c, \psi_t)      & 
        = \Omega_t \bm R (\bm \theta_c, \psi_t) \tilde{\bm U}_{a} \, , \label{eq_ua}      \\
  \bm \Omega_a (\bm \theta_c, \psi_t) & 
        = \Omega_t \bm R (\bm \theta_c, \psi_t) \tilde{\bm \Omega}_{a} \, , \label{eq_wa}
\end{align}
where the rotation matrix $\bm R$ 
(that transforms the expression of a vector
in the body reference frame into its expression in the 
global reference frame) is a function of $\theta$, $\phi$ and $\psi'$
%%%%%%%%%%%%%%%%%%%%%%%%%%%%%%%%%%%%%%%%%%%%%%%%%%%%%%%%%%%%%%%%
\begin{align}
  \bm R & =
  \begin{bmatrix}
    \mathrm{C}\phi \mathrm{C}\psi'   \mathrm{C}\theta - \mathrm{S}\phi   \mathrm{S} \psi'
     & - \mathrm{C}\psi' \mathrm{S}\phi - \mathrm{C}\phi    \mathrm{C}\theta \mathrm{S} \psi'
     & \mathrm{C}\phi \mathrm{S}\theta                                                        \\
    \mathrm{C}\psi' \mathrm{C}\theta \mathrm{S}\phi + \mathrm{C}\phi \mathrm{S}\psi'
     & \mathrm{C}\phi \mathrm{C}\psi' - \mathrm{C}\theta \mathrm{S}\phi \mathrm{S}\psi'
     & \mathrm{S}\phi \mathrm{S}\theta                                                        \\
    - \mathrm{C}\psi' \mathrm{S}\theta
     & \mathrm{S}\psi' \mathrm{S}\theta
     & \mathrm{C}\theta
  \end{bmatrix} \, , \label{eq_rotM}
\end{align}
%%%%%%%%%%%%%%%%%%%%%%%%%%%%%%%%%%%%%%%%%%%%%%%%%%%%%%%%%%%%%%%%
\tb{where $\mathrm{S}\theta$ stands for $\sin{(\theta)}$ 
and $\mathrm{C}\theta$ stands for $\cos{(\theta)}$
and so on.}

{\sl Passive motion.} Now, we discuss the passive part $(\bm U_p, \bm \Omega_p)$ 
induced by the external flow $\bm u_b$. This is a quite classical problem that we briefly
revise for completeness~\cite{kim1991microhydrodynamics,happel2012low}.
Taylor expansion allows to locally decompose the generic flow field $\bm u_b$  
into three parts,
\begin{align}
  u_i^b(\bm x_c + \delta \bm x_c) & = u_i^b(\bm x_c) + E_{ij}(\bm x_c) \delta x_j^c + S_{ij}(\bm x_c) \delta x_j^c \, , \label{eq_ub_taylor} \\
  E_{ij}(\bm x_c)                 & = \dfrac{1}{2}(u_{i,j}^b(\bm x_c) + u_{j,i}^b(\bm x_c)) \, ,                                             \\
  S_{ij}(\bm x_c)                 & = \dfrac{1}{2}(u_{i,j}^b(\bm x_c) - u_{j,i}^b(\bm x_c)) \, ,
\end{align}
where $E_{ij}$ and $S_{ij}$ are the symmetric and asymmetric part of the velocity gradient
$u_{i,j} = \partial u_i/\partial x_j$.
The first term of the right hand side of~\eqref{eq_ub_taylor} gives a pure rigid body 
translation $\bm U_p^b(\bm x_c)$ of the microswimmer without rotation, see~\figref{fig_apx_decoupling}(b). 
Instead, the effect of the last term induced a pure rigid body rotation 
$\bm \Omega_p^{S} = \frac{1}{2} \nabla \times \bm u^b$ where the $\nabla \times \bm u^b$ 
is the bulk fluid vorticity, see \figref{fig_apx_decoupling} (d).
\begin{table}%[htb]
  \setlength{\tabcolsep}{2mm}
  \renewcommand\arraystretch{1.1}
  \centering{
    \begin{tabular}{c c c}
      \toprule
      $k$ & strain rate base $\tilde E_{ij}^k$ & associated flow $\tilde u_i^{Ek}$  \\
      \midrule
      \addlinespace[1ex]
      $1$ & $ 
      \begin{bmatrix}
          1 & 0  & 0 \\
          0 & -1 & 0 \\
          0 & 0  & 0
      \end{bmatrix}$ &
      $(x_1, -x_2, 0)$  \\
      $2$ & $ 
      \begin{bmatrix}
          0 & 0  & 0 \\
          0 & -1 & 0 \\
          0 & 0  & 1
      \end{bmatrix}$ &
      $(0, -x_2, x_3)$ \\
      $3$ & $ 
      \begin{bmatrix}
          0 & 1 & 0 \\
          1 & 0 & 0 \\
          0 & 0 & 0
      \end{bmatrix}$ &
      $(x_2, x_1, 0)$  \\
      $4$ & $ 
      \begin{bmatrix}
          0 & 0 & 1 \\
          0 & 0 & 0 \\
          1 & 0 & 0
      \end{bmatrix}$ &
      $(x_3, 0, x_1)$  \\
      $5$ & $ 
      \begin{bmatrix}
          0 & 0 & 0 \\
          0 & 0 & 1 \\
          0 & 1 & 0
      \end{bmatrix}$ &
      $(0, x_3, x_2)$  \\
      \addlinespace[1ex]
      \bottomrule
    \end{tabular}}
  \caption{Base flow associated with the decomposition of the symmetric component of the velocity gradient, 
           see~\eqref{eq_Eijk}}
  \label{tab_strain_base}
\end{table}
The contribution of the symmetric part of the gradient to the motion, \figref{fig_apx_decoupling}(c), 
however, is more complex. 
$E_{ij}$ has nine components, but since it is symmetric, i.e. $E_{ij}=E_{ji}$, 
and the fluid is incompressible, i.e. $\mathrm{tr} (E_{ij}) = E_{ii} = 0$, 
only five of them are independent. 
Our approach is firstly to express the strain rate $E_{ij}$ in
the  body reference frame 
\begin{align}
  \tilde E_{ij} &= R^T E_{ij} R \, , 
  \label{eq_EijRt}
\end{align}
where $R$ is the rotation matrix \eqref{eq_rotM}.
Then, we decompose it in five basic modes due to the linearity of the Stokes equations~\cite{happel2012low, chwang1975hydromechanics}. 
\begin{align}
  \tilde E_{ij} &= 
 \sum_{k=1}^5 \tilde \beta_k \tilde E_{ij}^k \, , 
  \label{eq_Eijk}
\end{align}
Indeed, any $\tilde E_{ij}^k$ can be expressed as
\begin{align}
  \tilde E_{ij} 
  = 
  \begin{bmatrix}
  \beta_1 & \beta_3 & \beta_4 \\
  \beta_3 & -\beta_1-\beta_2 & \beta_5 \\
  \beta_4 & \beta_5 & \beta_2
  \end{bmatrix} \, , 
\end{align}
by using the $5$ components reported in~\tabref{tab_strain_base}.
Given this decomposition, 
we numerically solve the swimming kinematics $(\tilde{\bm U}_k^{E}, \tilde{\bm \Omega}_k^{E})$ 
of the passive microswimmer for the five components and sum them with proper weights $\tilde \beta_k$
\begin{align}
  \tilde{\bm U}_p^{E}(\bm x, \bm \theta_c, \psi_t)      & 
        = \sum_{k=1}^5 \tilde \beta_k(\bm x, \bm \theta_c, \psi_t) \tilde{\bm U}_k^{E}      \, , \\
  \tilde{\bm \Omega}_p^{E}(\bm x, \bm \theta_c, \psi_t) & 
        = \sum_{k=1}^5 \tilde \beta_k(\bm x, \bm \theta_c, \psi_t) \tilde{\bm \Omega}_k^{E} \, .
\end{align}
Finally, we express $\tilde{\bm U}_p^{E}$ and $\tilde{\bm \Omega}_p^{E}$
in the global reference frame
\begin{align}
  \bm U_p^E(\bm x, \bm \theta_c, \psi_t)              & 
        = \bm R (\bm \theta_c, \psi_t)\tilde{\bm U}_k^{E}(\bm x, \bm \theta_c, \psi_t)      \, , \\        
  \bm \Omega_p^E(\bm x, \bm \theta_c, \psi_t)         & 
        = \bm R (\bm \theta_c, \psi_t)\tilde{\bm \Omega}_k^{E}(\bm x, \bm \theta_c, \psi_t) \, . 
\end{align}

It is worth noting that the weights $\tilde \beta_k, k=1 \dots 5$ are functions of external 
flow $\bm u_b$ and swimmer configuration $(\bm \theta_c, \psi_t)$ and they 
do not vary with the geometric details of the microswimmer. 
Similar strategies for calculating the passive motion of the microswimmer 
can be found in~\cite{marath2018inertial, subramanian2006inertial}.

In summary, the microswimmer generalized velocity in an external flow is
obtained as
\begin{align}
  \bm U      & = 
        \Omega_t \bm R  \tilde{\bm U}_{a} + \bm U_p^b + 
        \bm R \sum_{k=1}^5 \tilde \beta_k \tilde{\bm U}_k^{E}      \, , \label{eq_base_u_apx}        \\
  \bm \Omega & = 
        \Omega_t \bm R  \tilde{\bm \Omega}_{a} + \bm \Omega_p^S + 
        \bm R \sum_{k=1}^5 \tilde \beta_k \tilde{\bm \Omega}_k^{E} \, . \label{eq_base_w_apx}
\end{align}
A sketch of the proposed  decoupling is reported in~\figref{fig_apx_decoupling}.
The main advantage of this method is that, for given geometry of microswimmer, 
regardless the tail spin rate $\Omega_t$, only six simulations are necessary;
one for getting $(\tilde{\bm U}_{a}, \tilde{\bm \Omega}_{a})$ and five 
for $(\tilde{\bm u}_k^{E}, \tilde{\bm \Omega}_k^{E}), k=1 \dots 5$. 
Thus, one can obtain these quantities accurately previously, 
and then solve the microswimmer kinematics~\eqref{eq_kinematic1}-\eqref{eq_kinematic3}.

For the microswimmer motion in the shear flow $\bm u_b=(X_3, 0, 0)$, we have
\begin{align}
  E_{ij} &= \dfrac{1}{2}
  \begin{bmatrix}
    0 & 0 & 1 \\
    0 & 0 & 0 \\
    1 & 0 & 0 
  \end{bmatrix} \, ,
\end{align}
that, using~\eqref{eq_EijRt} and~\eqref{eq_Eijk}, gives 
\begin{align}
  \tilde \beta_1 & = 
      \mathrm{C} \psi' (\mathrm{S} \phi \mathrm{S} \psi'-\mathrm{C} \phi \mathrm{C} \psi' \mathrm{C} \theta) \mathrm{S} \theta \, ,           \\
  \tilde \beta_2 & = 
      \mathrm{C} \phi \mathrm{C} \theta \mathrm{S} \theta \, ,                                                                                \\
  \tilde \beta_3 & = 
      \dfrac{1}{4} (2 \mathrm{C} (2 \psi') \mathrm{S} \phi \mathrm{S} \theta+\mathrm{C} \phi \mathrm{S} (2 \psi') \mathrm{S} (2 \theta)) \, , \\
  \tilde \beta_4 & = 
      \dfrac{1}{2} (\mathrm{C} \phi \mathrm{C} \psi' \mathrm{C} (2 \theta)-\mathrm{C} \theta \mathrm{S} \phi \mathrm{S} \psi') \, ,           \\
  \tilde \beta_5 & = 
      \dfrac{1}{2} (-\mathrm{C} \psi' \mathrm{C} \theta \mathrm{S} \phi-\mathrm{C} \phi \mathrm{C} (2 \theta) \mathrm{S} \psi') \, .
\end{align}
%

%%%%%%%%%%%%%%%%%%%%%%%%%%%%%%%%%%%%%%%%%%%%%%%%%%%%%%%%%%%%%%%%%%%
\begin{figure}[htb]
  \includegraphics[width=\columnwidth]{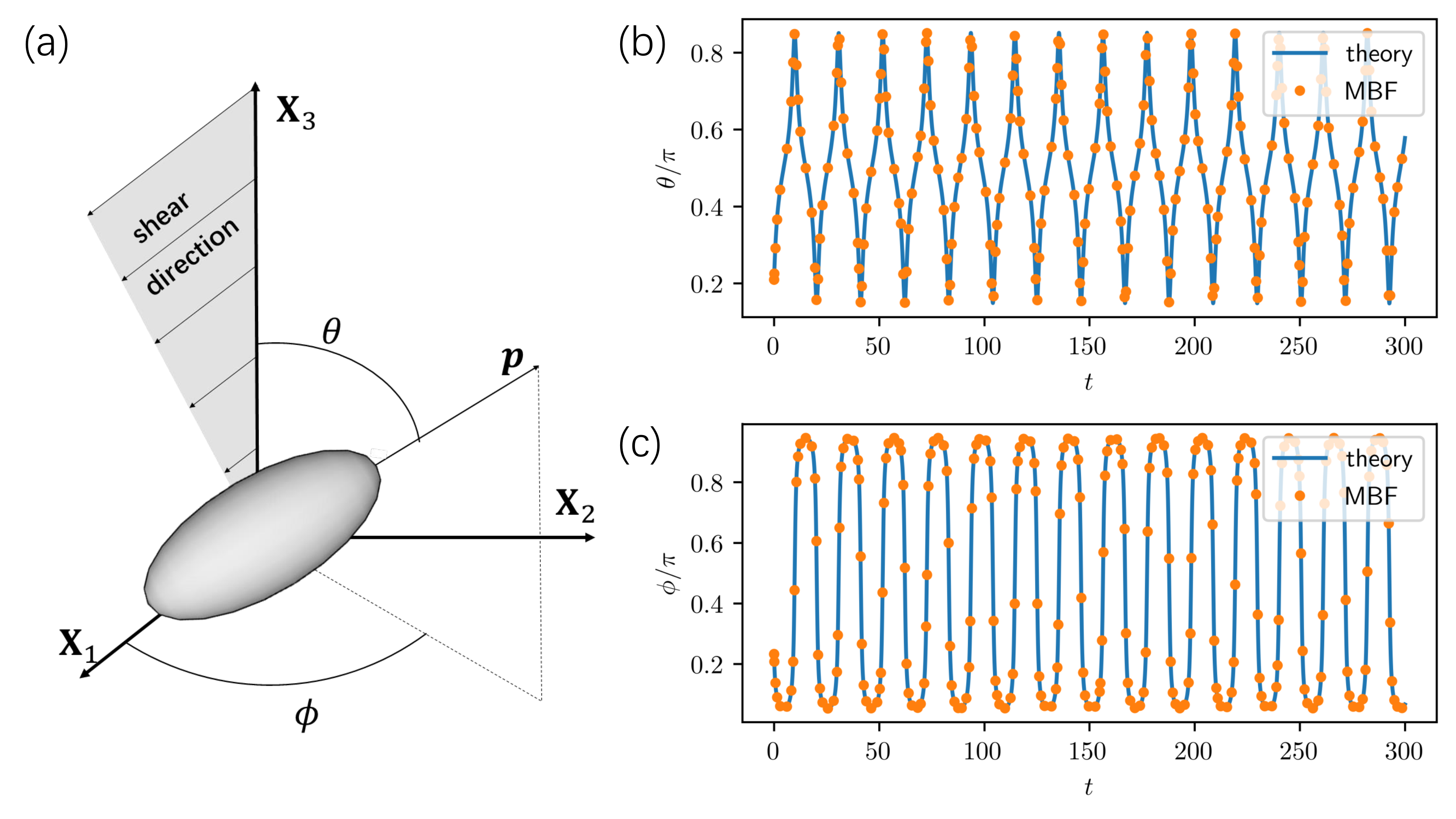}
  \caption{
   Validation of the numerical method: Jeffery orbits.  
   (a) Sketch of an ellipse orbit in a shear flow. 
   The shear flow is in the $\textbf{X}_1\textbf{X}_3$ plane of the global coordinate frame 
   $\textbf{O}\textbf{X}_1\textbf{X}_2\textbf{X}_3$. 
   The polar $\theta$ and azimuthal $\phi$ angles 
   are used to describe the orientation of the body frame with respect to the global frame.
   The unit vector $\bm{p}$ denotes the orientation of the ellipsoid. 
   (b, c) Time evolution of angles $\theta$ and $\phi$ for an ellipse with aspect ratio $r_{h1} / r_{h2} = 3$ moving in a shear flow. 
   The initial orientation of the ellipse is $(\theta = 0.21 \pi, \phi = 0.23 \pi)$. 
Orange points 
   represent our numerical solution while the analytical solution~\cite{kim1991microhydrodynamics}
are reported as blue lines.} 
  \label{fig_apx_varify}
\end{figure}
%%%%%%%%%%%%%%%%%%%%%%%%%%%%%%%%%%%%%%%%%%%%%%%%%%%%%%%%%%%%%%%%%%%

To test our approach, we reproduced the Jeffery orbit \cite{jeffery1922motion} for a $r_{h1} / r_{h2} = 3$ ellipse in a shear flow, see fig~\ref{fig_apx_varify}.

%%%%%%%%%%%%%%%%%%%%%%%%%%%%%%%%%%%%%%%%%%%%%%%%%%%%%%%%%%%%%%%%%%
\bibliographystyle{unsrt}
%\bibliography{jizhang2020_ref.bib}

\begin{thebibliography}{1}

\bibitem{Note1}
See Supplemental Material at [URL will be inserted by publisher] for movies
  showing these trajectories.

\bibitem{Note2}
See Supplemental Material at [URL will be inserted by publisher] for a figure
  representing the dependence of the final stable trajectory on initial swimmer
  orientation.

\bibitem{Note3}
Considering one-quarter of an ellipse, the arc length is a monotone increasing
  function of the eccentric angle $\xi $ that has no explicit expression.
  Therefore, we first fit this function using a quadratic polynomial, then
  determine a set of $\xi _i, i \in (1, n_h)$ that keeps the distance between
  two adjacent points approximately equal. Finally we calculate the location of
  the points on the $\protect \textbf {e}_3 \protect \textbf {e}_1$ plane,
  see~\cite {Andy2020}.

\end{thebibliography}


\begin{thebibliography}{10}

\bibitem{berg2008coli}
Howard~C Berg.
\newblock {\em E. coli in Motion}.
\newblock Springer Science \& Business Media, 2008.

\bibitem{qian2013bacterial}
Chen Qian, Chui~Ching Wong, Sanjay Swarup, and Keng-Hwee Chiam.
\newblock Bacterial tethering analysis reveals a “run-reverse-turn”
  mechanism for pseudomonas species motility.
\newblock {\em Applied and environmental microbiology}, 79(15):4734--4743,
  2013.

\bibitem{sartori2018wall}
Paolo Sartori, Enrico Chiarello, Gaurav Jayaswal, Matteo Pierno, Giampaolo
  Mistura, Paola Brun, Adriano Tiribocchi, and Enzo Orlandini.
\newblock Wall accumulation of bacteria with different motility patterns.
\newblock {\em Physical Review E}, 97(2):022610, 2018.

\bibitem{zhang2010artificial}
Li~Zhang, Kathrin~E Peyer, and Bradley~J Nelson.
\newblock Artificial bacterial flagella for micromanipulation.
\newblock {\em Lab on a Chip}, 10(17):2203--2215, 2010.

\bibitem{mhanna2014artificial}
Rami Mhanna, Famin Qiu, Li~Zhang, Yun Ding, Kaori Sugihara, Marcy Zenobi-Wong,
  and Bradley~J Nelson.
\newblock Artificial bacterial flagella for remote-controlled targeted
  single-cell drug delivery.
\newblock {\em Small}, 10(10):1953--1957, 2014.

\bibitem{lauga2009hydrodynamics}
Eric Lauga and Thomas~R Powers.
\newblock The hydrodynamics of swimming microorganisms.
\newblock {\em Reports on Progress in Physics}, 72(9):096601, 2009.

\bibitem{elgeti2015physics}
Jens Elgeti, Roland~G Winkler, and Gerhard Gompper.
\newblock Physics of microswimmers—single particle motion and collective
  behavior: a review.
\newblock {\em Reports on progress in physics}, 78(5):056601, 2015.

\bibitem{lauga2006swimming}
Eric Lauga, Willow~R DiLuzio, George~M Whitesides, and Howard~A Stone.
\newblock Swimming in circles: motion of bacteria near solid boundaries.
\newblock {\em Biophysical journal}, 90(2):400--412, 2006.

\bibitem{guccione2017diffusivity}
Giorgia Guccione, Daniela Pimponi, Paolo Gualtieri, and Mauro Chinappi.
\newblock Diffusivity of e. coli-like microswimmers in confined geometries: The
  role of the tumbling rate.
\newblock {\em Physical Review E}, 96(4):042603, 2017.

\bibitem{shum2010modelling}
H~Shum, EA~Gaffney, and DJ~Smith.
\newblock Modelling bacterial behaviour close to a no-slip plane boundary: the
  influence of bacterial geometry.
\newblock {\em Proceedings of the Royal Society A: Mathematical, Physical and
  Engineering Sciences}, 466(2118):1725--1748, 2010.

\bibitem{di2011swimming}
R~Di~Leonardo, D~Dell’Arciprete, L~Angelani, and V~Iebba.
\newblock Swimming with an image.
\newblock {\em Physical review letters}, 106(3):038101, 2011.

\bibitem{pimponi2016hydrodynamics}
Daniela Pimponi, Mauro Chinappi, Paolo Gualtieri, and Carlo~Massimo Casciola.
\newblock Hydrodynamics of flagellated microswimmers near free-slip interfaces.
\newblock {\em Journal of Fluid Mechanics}, 789:514--533, 2016.

\bibitem{hu2015physical}
Jinglei Hu, Adam Wysocki, Roland~G Winkler, and Gerhard Gompper.
\newblock Physical sensing of surface properties by microswimmers--directing
  bacterial motion via wall slip.
\newblock {\em Scientific reports}, 5:9586, 2015.

\bibitem{bianchi20193d}
Silvio Bianchi, Filippo Saglimbeni, Giacomo Frangipane, Dario Dell'Arciprete,
  and Roberto Di~Leonardo.
\newblock 3d dynamics of bacteria wall entrapment at a water--air interface.
\newblock {\em Soft matter}, 15(16):3397--3406, 2019.

\bibitem{fu2012bacterial}
Henry~C Fu, Thomas~R Powers, and Roman Stocker.
\newblock Bacterial rheotaxis.
\newblock {\em Proceedings of the National Academy of Sciences},
  109(13):4780--4785, 2012.

\bibitem{fu2009separation}
Marcos, Henry~C. Fu, Thomas~R. Powers, and Roman Stocker.
\newblock Separation of microscale chiral objects by shear flow.
\newblock {\em Phys. Rev. Lett.}, 102:158103, Apr 2009.

\bibitem{rusconi2014bacterial}
Roberto Rusconi, Jeffrey~S Guasto, and Roman Stocker.
\newblock Bacterial transport suppressed by fluid shear.
\newblock {\em Nature physics}, 10(3):212, 2014.

\bibitem{kanehl2014fluid}
Philipp Kanehl and Takuji Ishikawa.
\newblock Fluid mechanics of swimming bacteria with multiple flagella.
\newblock {\em Physical Review E}, 89(4):042704, 2014.

\bibitem{riley2018swimming}
Emily~E Riley, Debasish Das, and Eric Lauga.
\newblock Swimming of peritrichous bacteria is enabled by an elastohydrodynamic
  instability.
\newblock {\em Scientific reports}, 8, 2018.

\bibitem{einarsson2016tumbling}
J~Einarsson, BM~Mihiretie, A~Laas, S~Ankardal, JR~Angilella, D~Hanstorp, and
  B~Mehlig.
\newblock Tumbling of asymmetric microrods in a microchannel flow.
\newblock {\em Physics of Fluids}, 28(1):013302, 2016.

\bibitem{einarsson2015angular}
Jonas Einarsson.
\newblock {\em Angular dynamics of small particles in fluids}.
\newblock PhD thesis, Department of Physics, University of Gothenburg, 2015.

\bibitem{thorp2019motion}
Ian Thorp and John Lister.
\newblock Motion of a non-axisymmetric particle in viscous shear flow.
\newblock {\em Journal of Fluid Mechanics}, 2019.

\bibitem{berg1993torque}
Howard~C Berg and Linda Turner.
\newblock Torque generated by the flagellar motor of escherichia coli.
\newblock {\em Biophysical journal}, 65(5):2201--2216, 1993.

\bibitem{xing2006torque}
Jianhua Xing, Fan Bai, Richard Berry, and George Oster.
\newblock Torque--speed relationship of the bacterial flagellar motor.
\newblock {\em Proceedings of the National Academy of Sciences},
  103(5):1260--1265, 2006.

\bibitem{young2006method}
DL~Young, SJ~Jane, CM~Fan, K~Murugesan, and CC~Tsai.
\newblock The method of fundamental solutions for 2d and 3d stokes problems.
\newblock {\em Journal of Computational Physics}, 211(1):1--8, 2006.

\bibitem{Note1}
See Supplemental Material at [URL will be inserted by publisher] for movies
  showing these trajectories.

\bibitem{Note2}
See Supplemental Material at [URL will be inserted by publisher] for a figure
  representing the dependence of the final stable trajectory on initial swimmer
  orientation.

\bibitem{berg2000motile}
Howard Berg.
\newblock Motile behavior of bacteria.
\newblock {\em Physics today}, 2000.

\bibitem{liu2014propulsion}
Bin Liu, Kenneth~S Breuer, and Thomas~R Powers.
\newblock Propulsion by a helical flagellum in a capillary tube.
\newblock {\em Physics of Fluids}, 26(1):011701, 2014.

\bibitem{rorai2019limitations}
C~Rorai, M~Zaitsev, and S~Karabasov.
\newblock On the limitations of some popular numerical models of flagellated
  microswimmers: importance of long-range forces and flagellum waveform.
\newblock {\em Royal Society open science}, 6(1):180745, 2019.

\bibitem{cortez2005method}
Ricardo Cortez, Lisa Fauci, and Alexei Medovikov.
\newblock The method of regularized stokeslets in three dimensions: analysis,
  validation, and application to helical swimming.
\newblock {\em Physics of Fluids}, 17(3):031504, 2005.

\bibitem{zhang2020active}
Bokai Zhang, Yang Ding, and Xinliang Xu.
\newblock Active suspensions of bacteria and passive objects: a model for the
  near field pair dynamics.
\newblock {\em arXiv preprint arXiv:2002.04693}, 2020.

\bibitem{klaseboer2012non}
Evert Klaseboer, Qiang Sun, and Derek~YC Chan.
\newblock Non-singular boundary integral methods for fluid mechanics
  applications.
\newblock {\em Journal of Fluid Mechanics}, 696:468--478, 2012.

\bibitem{muldowney1995spectral}
GP~Muldowney and Jonathan~JL Higdon.
\newblock A spectral boundary element approach to three-dimensional stokes
  flow.
\newblock {\em Journal of Fluid Mechanics}, 298:167--192, 1995.

\bibitem{mathijssen2019oscillatory}
Arnold~JTM Mathijssen, Nuris Figueroa-Morales, Gaspard Junot, {\'E}ric
  Cl{\'e}ment, Anke Lindner, and Andreas Z{\"o}ttl.
\newblock Oscillatory surface rheotaxis of swimming e. coli bacteria.
\newblock {\em Nature communications}, 10(1):1--12, 2019.

\bibitem{kim1991microhydrodynamics}
Sangtae Kim and Seppo~J Karrila.
\newblock {\em Microhydrodynamics: principles and selected applications}.
\newblock Courier Corporation, 1991.

\bibitem{najafi2004simple}
Ali Najafi and Ramin Golestanian.
\newblock Simple swimmer at low reynolds number: Three linked spheres.
\newblock {\em Physical Review E}, 69(6):062901, 2004.

\bibitem{daddi2020tuning}
Abdallah Daddi-Moussa-Ider, Maciej Lisicki, and Arnold~JTM Mathijssen.
\newblock Tuning the upstream swimming of microrobots by shape and cargo size.
\newblock {\em Physical Review Applied}, 14(2):024071, 2020.

\bibitem{aboelkassem2013stokeslets}
Yasser Aboelkassem and Anne~E Staples.
\newblock Stokeslets-meshfree computations and theory for flow in a collapsible
  microchannel.
\newblock {\em Theoretical and Computational Fluid Dynamics}, 27(5):681--700,
  2013.

\bibitem{lockerby2016fundamental}
Duncan~A Lockerby and B~Collyer.
\newblock Fundamental solutions to moment equations for the simulation of
  microscale gas flows.
\newblock {\em Journal of Fluid Mechanics}, 806:413--436, 2016.

\bibitem{Note3}
Considering one-quarter of an ellipse, the arc length is a monotone increasing
  function of the eccentric angle $\xi $ that has no explicit expression.
  Therefore, we first fit this function using a quadratic polynomial, then
  determine a set of $\xi _i, i \in (1, n_h)$ that keeps the distance between
  two adjacent points approximately equal. Finally we calculate the location of
  the points on the $\protect \textbf {e}_3 \protect \textbf {e}_1$ plane,
  see~\cite {Andy2020}.

\bibitem{Andy2020}
Andy Jones.
\newblock {How to divide an ellipse to equal segments?}
\newblock
  \url{https://stackoverflow.com/questions/20197974/how-to-divide-an-ellipse-to-equal-segments},
  2013.
\newblock [Online; accessed 6-August-2020].

\bibitem{saad1986gmres}
Youcef Saad and Martin~H Schultz.
\newblock Gmres: A generalized minimal residual algorithm for solving
  nonsymmetric linear systems.
\newblock {\em SIAM Journal on scientific and statistical computing},
  7(3):856--869, 1986.

\bibitem{balay1997efficient}
Satish Balay, William~D Gropp, Lois~Curfman McInnes, and Barry~F Smith.
\newblock Efficient management of parallelism in object-oriented numerical
  software libraries.
\newblock In {\em Modern software tools for scientific computing}, pages
  163--202. Springer, 1997.

\bibitem{petsc-web-page}
Satish Balay, Shrirang Abhyankar, Mark~F. Adams, Jed Brown, Peter Brune, Kris
  Buschelman, Lisandro Dalcin, Alp Dener, Victor Eijkhout, William~D. Gropp,
  Dmitry Karpeyev, Dinesh Kaushik, Matthew~G. Knepley, Dave~A. May,
  Lois~Curfman McInnes, Richard~Tran Mills, Todd Munson, Karl Rupp, Patrick
  Sanan, Barry~F. Smith, Stefano Zampini, Hong Zhang, and Hong Zhang.
\newblock {PETS}c {W}eb page.
\newblock \url{https://www.mcs.anl.gov/petsc}, 2019.

\bibitem{pimponi2018flagellated}
Daniela Pimponi, Mauro Chinappi, and Paolo Gualtieri.
\newblock Flagellated microswimmers: Hydrodynamics in thin liquid films.
\newblock {\em The European Physical Journal E}, 41(2):1--8, 2018.

\bibitem{graf2008quaternions}
Basile Graf.
\newblock Quaternions and dynamics.
\newblock {\em arXiv preprint arXiv:0811.2889}, 2008.

\bibitem{diebel2006representing}
James Diebel.
\newblock Representing attitude: Euler angles, unit quaternions, and rotation
  vectors.
\newblock {\em Matrix}, 58(15-16):1--35, 2006.

\bibitem{bogacki1996efficient}
P~Bogacki and Lawrence~F Shampine.
\newblock An efficient runge-kutta (4, 5) pair.
\newblock {\em Computers \& Mathematics with Applications}, 32(6):15--28, 1996.

\bibitem{abhyankar2018petsc}
Shrirang Abhyankar, Jed Brown, Emil~M Constantinescu, Debojyoti Ghosh, Barry~F
  Smith, and Hong Zhang.
\newblock Petsc/ts: A modern scalable ode/dae solver library.
\newblock {\em arXiv preprint arXiv:1806.01437}, 2018.

\bibitem{happel2012low}
John Happel and Howard Brenner.
\newblock {\em Low Reynolds number hydrodynamics: with special applications to
  particulate media}, volume~1.
\newblock Springer Science \& Business Media, 2012.

\bibitem{chwang1975hydromechanics}
Allen~T Chwang and T~Wu.
\newblock Hydromechanics of low-reynolds-number flow. part 2. singularity
  method for stokes flows.
\newblock {\em Journal of Fluid mechanics}, 67(4):787--815, 1975.

\bibitem{subramanian2006inertial}
G~Subramanian and DL~Koch.
\newblock Inertial effects on the orientation of nearly spherical particles in
  simple shear flow.
\newblock {\em Journal of Fluid Mechanics}, 557:257, 2006.

\bibitem{banerjee2020anisotropic}
Mahan~Raj Banerjee and Ganesh Subramanian.
\newblock An anisotropic particle in a simple shear flow: an instance of
  chaotic scattering.
\newblock {\em arXiv preprint arXiv:2005.11157}, 2020.

\bibitem{marath2018inertial}
Navaneeth~K Marath and Ganesh Subramanian.
\newblock The inertial orientation dynamics of anisotropic particles in planar
  linear flows.
\newblock {\em Journal of Fluid Mechanics}, 844:357, 2018.

\bibitem{jeffery1922motion}
George~Barker Jeffery.
\newblock The motion of ellipsoidal particles immersed in a viscous fluid.
\newblock {\em Proceedings of the Royal Society of London. Series A, Containing
  papers of a mathematical and physical character}, 102(715):161--179, 1922.

\end{thebibliography}

\end{document}